\documentclass{aastex63}

\usepackage{xcolor}

\submitjournal{ApJ}

\begin{document}

\title{Statistical Analysis of the Relation between Coronal Mass Ejections and Solar Energetic Particles}

\author{Kosuke Kihara}
\affiliation{Astronomical Observatory, Kyoto University, Sakyo, Kyoto 606-8502, Japan}

\author{Yuwei Huang}
\affiliation{Astronomical Observatory, Kyoto University, Sakyo, Kyoto 606-8502, Japan}

\author{Nobuhiko Nishimura}
\affiliation{Institute for Space-Earth Environmental Research, Nagoya University, Nagoya, Japan}

\author{Nariaki V. Nitta}
\affiliation{Lockheed Martin Solar and Astrophysics Laboratory, Palo Alto, CA 94304 USA}

\author{Seiji Yashiro}
\affiliation{Department of Physics, The Catholic University of America, Washington, USA}

\author{Kiyoshi Ichimoto}
\affiliation{Astronomical Observatory, Kyoto University, Sakyo, Kyoto 606-8502, Japan}

\author{Ayumi Asai}
\affiliation{Astronomical Observatory, Kyoto University, Sakyo, Kyoto 606-8502, Japan}



\begin{abstract}

To improve the forecasting capability of impactful solar energetic particle (SEP) events, the relation between coronal mass ejections (CMEs) and SEP events needs to be better understood. Here we present a statistical study of SEP occurrences and timescales with respect to the CME source locations and speeds, considering all 257 fast ($v_{CME} \geq$ 900 km s$^{-1}$) and wide (angular width $\geq$ 60$\arcdeg$) CMEs that occurred between December 2006 and October 2017. We associate them with SEP events at energies above 10~MeV. Examination of the source region of each CME reveals that CMEs more often accompany a SEP event if they originate from the longitude of E20\,--\,W100 relative to the observer. However, a SEP event could still be absent if the CME is $<$ 2000~km~s$^{-1}$. For the associated CME-SEP pairs, we compute three timescales for each of the SEP events, following \citet{kahler_2005_apj, kahler_2013_apj}; namely the timescale of the onset (TO), the rise time (TR), and the duration (TD). They are correlated with the longitude of the CME source region relative to the footpoint of the Parker spiral ($\Delta\Phi$) and $v_{CME}$. The TO tends to be short for $| \Delta \Phi | < $60$\arcdeg$. This trend is weaker for TR and TD. The SEP timescales are only weakly correlated with $v_{CME}$. Positive correlations of both TR and TD with $v_{CME}$ are seen in poorly connected (large $| \Delta \Phi |$) events. Additionally, TO appears to be negatively correlated with $v_{CME}$ for events with small $| \Delta \Phi |$.

\end{abstract}

\keywords{coronal mass ejections (CMEs) --- Sun: particle emission}


\section{Introduction} \label{sec:intro}
Solar energetic particles (SEPs) may give rise to major space weather hazards.  The National Oceanic and Atmospheric Administration (NOAA) space weather scale for solar radiation storms characterizes the severity of the effects in three areas (biological, satellite operations, and other systems such as high-frequency communications) in accordance with the peak flux of $>$10~MeV ions\footnote{\url{https://www.swpc.noaa.gov/noaa-scales-explanation}}.  
For the practical purpose of preparing for possible space weather impacts, it is important to predict whether a SEP event will occur, when it will start, how intense it will become, and how long the SEP flux will stay above a given threshold.  Even though a number of schemes have been proposed for SEP forecasting 
\citep[e.g.,][and references therein]{anastasiadis_2017_sp},
we still cannot reliably predict SEP events even after the possibly associated solar activity phenomenon is observed. This is obviously because our science-based understanding of the origins of SEP events is far from sufficient.

We often classify SEP events into two groups, impulsive and gradual 
\citep[see][]{reames_1999_ssr, reames_2013_ssr}
on the basis of their observed properties, including timescales, spectra, composition and charge states, and the associated radio bursts. In this scheme, 
it is the gradual SEP events with high ion (mostly proton) fluxes that can cause hazardous space weather conditions. Their close association with coronal mass ejections (CMEs)---as shown, for example, by 
\citet{kahler_1978_sp, kahler_1984_jgr}
indicates that shock waves driven by energetic CMEs are responsible for energizing the ions in gradual SEP events, typically through diffusive shock acceleration 
\citep{lee_2012_ssr, desai_2016_lrsp}.

Before the discovery of CMEs in the early 1970s, solar flares were thought to play a central role in causing coronal and interplanetary disturbances including protons observed in situ
\citep[e.g.,][]{lin_1976_sp, gosling_1993_jgr}.
However, in the two-class paradigm \citep{reames_1999_ssr, reames_2013_ssr}, solar flares powered by magnetic reconnection are relevant primarily to impulsive SEP events that are typically enhanced in $^3$He and heavy ions such as Fe, as compared with the elemental composition of the solar wind.  Such compositional anomalies are hard to explain if the solar wind particles are accelerated by shock waves, as this requires some stochastic processes that can change the composition \citep[see, for example,][]{miller_1997_jgr}.

There has been renewed interest in the role of solar flares in producing large gradual SEP events. This is due to the intimate association of gradual SEP events with type III radio bursts \citep[which have been considered ``flare'' attributes, see][]{cane_2002_jgr} and to the apparent correlations between SEPs and flare parameters 
\citep{dierckxsens_2015_sp, grechnev_2015_sp, trottet_2015_sp}. However, these arguments may not exclude CMEs as the main contributor for gradual SEP events for the following reasons.  First, CMEs are also frequently accompanied by type III bursts. Second, the parameters of large flares may vary in proportion to CME parameters as a result of the so-called ``big-flare syndrome'' \citep{kahler_1982_jgr}. Moreover, there are no SEPs from intense flares if they are not associated with CMEs \cite[e.g., all the X-class flares in AR~12192 in October 2014, see][]{sun_2015_apj},
and some of the most intense SEP events can be associated with flares that are quite modest \citep{cliver_2016_apj}.

Therefore, we assume that particles in gradual SEP events, at least the large ones, are accelerated by CME-driven shock waves. In this assumption, we may expect a correlation between the SEP peak flux and the CME speed, which is generally the case, although for a given CME speed, a scatter of up to four orders of magnitude in the SEP peak fluxes was found \citep{kahler_2001_jgr}. This large scatter can be attributed to a number of factors, ranging from the conditions for particle acceleration at the CME-driven shocks to the transport processes undergone by the particles. Earlier events may set up preconditioning in favor of SEP production by providing seed particles and producing enhanced levels of turbulence at the shock \citep[e.g.,][]{li_2005_icrc}.  Observationally, a CME preceded by another CME within a short interval tends to be more SEP-productive \citep{gopalswamy_2004_jgr, kahler_2005_jgr}.  Additionally, even though the CME speed is a good measure of the shock speed, the efficiency of particle acceleration depends on various shock parameters that may vary significantly over the shock surface. The SEP flux may be affected by where on the shock the observer is dynamically connected to \citep[e.g.,][and refferences therein]{kouloumvakos_2019_apj}. These factors, together with the transport effects, that might involve cross-field diffusion \citep[e.g.,][]{zhang_2009_apj}, affect not only the measured SEP peak but also the SEP temporal variations. The latter may often be consistent with the patterns expected from the longitude of the source region relative to the observer \citep{cane_1988_jgr}, but occasionally SEP events with prompt onsets may be observed even from poorly connected longitudes \citep[e.g.,][]{cliver_1982_sp, gomez_herrero_2015_apj}. It is likely that the observed SEP peak fluxes and temporal variations result from a combination of the above-mentioned factors. With this in mind, it is meaningful to study SEP events statistically in relation to the CME speed and the longitude of the source region.

In this paper, we present a statistical study of SEP occurrences and timescales with respect to CME source locations and speeds. 
Here we start from fast and wide CMEs and relate them to SEP properties.  Most previous studies have started from SEP events and have then studied the properties of the associated CMEs and flares, ignoring CMEs not associated with SEP events. 
The recent study of 11 CMEs which did not produce SEP events by \citet{lario_2020_apj} may be an exception.
Our work focuses on the presence/absence and timescales of SEP events in association with individual CMEs, as presented in Sections 3 and 4, respectively.  These are preceded by a description of our event list (Section 2) and followed by a discussion of how to explain our findings (Section 5). We summarize our conclusions in Section 6.

\section{Event List} \label{sec:evl}
Our ultimate goal is to understand how the properties of CMEs may affect the properties of SEP events, such as their occurrence, peak fluxes and timescales.  To acknowledge the fact that some energetic CMEs, even from well-connected longitudes, produce no SEPs or that CMEs from poorly-connected regions produce SEP events that quickly rise to a peak, it is meaningful to study all those CMEs irrespective of their associated SEPs and then to investigate the reasons for the wide variety of SEP properties. This approach complements one that discusses the properties of only those CMEs that are associated with SEP events \citep[e.g.,][]{kahler_2001_jgr}.  

Our study is based on fast ($v_{CME} \geq$ 900~km~s$^{-1}$) and wide (angular width $\geq$~60$\arcdeg$) CMEs. In the first approximation, these CMEs may be considered to drive the shocks that are responsible for accelerating the protons observed at 1 AU, although the occurrence of a shock wave depends not only on the CME speed, but also on the conditions of the ambient solar wind.  We imposed the restriction on angular width in order to exclude narrow CMEs, which are typically associated with small impulsive SEP events \citep{kahler_2001_apj}.
We selected them from the CDAW SOHO LASCO CME catalog\footnote{\url{https://cdaw.gsfc.nasa.gov/CME\_list/}} \citep{yashiro_2004_jgr}, which is a complete manually-generated catalog of CMEs as observed by the Large Angle and Spectrometric Coronagraph Experiment \citep[LASCO:][]{Brueckner_1995_sp} on board the Solar and Heliospheric Observatory (SOHO). Measurements of the kinematic parameters of CMEs included in the catalog come from visual inspection of all the available difference images.  

Another important factor that can affect the properties of SEP events is the magnetic field connection between the observer and the CME-driven shock wave, which may be assumed to expand concentrically from the source region of the CME. If the region is on the visible side of the Sun, we can locate it using known low coronal signatures of CMEs, such as coronal dimming and post-eruption arcades \citep[e.g.,][]{zhang_2007_jgr, hudson_2011_jgr, nitta_2014_sp}. These signatures are found in coronal images at extreme-ultraviolet (EUV) wavelengths. In order to maximize the number of CMEs for which source regions can be identified, including those from the far side, we 
have studied those CMEs that occurred since December 2006, 
so that we can make use of information from the EUV imagers on board the Solar-Terrestrial Relations Observatory (STEREO), in addition to those near the Sun-Earth line from SOHO (until 2010) and the Solar Dynamics Observatory \citep[SDO:][from 2010]{pesnell_2012_sp}. 
All the fast and wide CMEs in solar cycle 24 were included in the period of our investigation (i.e., from December 2006 to October 2017).
After examining the EUV images taken around the times of the 257 CMEs that meet our criteria for speed and angular width, we removed 18 CMEs for which source regions could not be identified.  Almost all of them occurred while no STEREO data were available around the great conjunction in 2015. 

Instead of discussing common SEP events observed by multiple spacecraft at separate longitudes \citep[e.g.,][]{richardson_2014_sp} or CMEs without SEPs at any of these spacecraft \citep{lario_2020_apj}, we studied the SEP events (or lack thereof) at Earth, STEREO-A, and STEREO-B that are associated with each of the 239 CMEs.  We thus have a total of 717 potential measurements. 
We extracted the time profiles of $>$10~MeV proton fluxes by using 
data with five-minute temporal resolution from the Energetic Particle Sensor \citep[EPS:][]{onsager_1996_spie} on the Geostationary
Operations Environmental Satellite (GOES) and from the High-Energy Telescope \citep[HET:][]{von_rosenvinge_2008_ssr} and the Low-Energy Telescope \citep[LET:][]{mewaldt_2008_ssr}, which belong to the suite of instruments for the In Situ Measurements of Particles and CME Transients \citep[IMPACT:][]{luhmann_2008_ssr} on STEREO. The $>$10~MeV integral flux is one of the standard products of GOES, but for STEREO / IMPACT we had to compute it by combining the HET and LET data, as illustrated by \citet{gopalswamy_2016_apj}.

Now we show how the $>$10~MeV proton flux compares between GOES and STEREO.  \citet{rodriguez_2017_sw} performed a cross-calibration between GOES and STEREO using two SEP events that occurred in 2006 December while STEREO-A and STEREO-B were still located near Earth. They reported that the STEREO 10\,--\,100 MeV flux were smaller than GOES: the first, the second, and the third quartile of STEREO-A (STEREO-B) to GOES ratios are 0.850 (0.874), 0.926 (0.948), and 1.017 (1.071), respectively. We carried out the same analysis for the $>$10~MeV integral flux and confirmed that the first, the second, and the third quartile are 0.841, 0.921, and 1.008 for STEREO-A, and 0.858, 0.935, and 1.042 for STEREO-B. The 5th and 95th percentiles of the ratios are 0.736 and 1.215 for STEREO-A, and 0.767 and 1.272 for STEREO-B. Therefore 90\% of $>$10~MeV integral fluxes agree within 27\%.

In Table~{\ref{allcme}}, we list the selected CMEs.  The first three columns show the onset date and time, the projected speed, and the angular width of the CME, taken from the CDAW SOHO LASCO CME catalog.  Note that the CME onset time is calculated by extrapolating the height-time profile in the LASCO field of view to the solar surface 
(1 solar radius from the Sun center).
The next two columns show the magnitude and location of the associated solar flare.  
The remaining columns show the quality and the peak proton flux, if it exceeded 1 particle flux unit (pfu: defined as particles s$^{-1}$ sr$^{-1}$ cm$^{-2}$), separately for GOES, STEREO-A, and STEREO-B. 
The ``quality'' is one of the following.

\begin{itemize}

\item Good (110 events): SEPs are detected unambiguously with the peak $>$10~MeV proton flux exceeding 1~pfu.  It is $>$10~pfu in 69 events and $\leq$10~pfu in 41 events. They are sufficiently well-observed that we can compute all the timescales (see Section~4). 

\item Contaminated (26 events): The observed SEP onset is clearly associated with the CME, but the later, post-peak temporal variations are contaminated by another SEP event due to a later CME or an energetic storm particle (ESP) event due to the shock wave driven by the present or an earlier CME\footnote{There are a few confusing cases, where an ESP-related shock arrived during the SEP event in question. We identified a non-ESP peak before the shock arrival as the SEP peak.}. 
In 15 events, the peak $>$10~MeV proton flux exceeds 10~pfu.  We were able to measure all the timescales but the duration (see Section~4). 

\item No SEP (395 events): No $>$10~MeV protons are detected exceeding the 1~pfu level during the normal background. The normal background is about 0.2~pfu for GOES and about 0.1~pfu for STEREO-A and STEREO-B. In 71 of them, we noted a smaller enhancement ($\leq$1~pfu), and they are so indicated.  

\item HiB$\_$N (85 events): The background was elevated from the normal due to earlier events, preventing a small SEP event from being detected. Despite the higher background, however, we clearly find the presence of a SEP event in 18 cases, of which 16 have the peak $>$10~MeV proton flux that exceeds 10~pfu. An approximate background level (in pfu) is indicated by $N$. Subsets of these events, depending on $N$, are excluded in the discussion of the SEP association rate of CMEs (see Section 3).  
 
\item No data (79 events): No SEP data are available around the time of the CME. This includes the periods of no STEREO data due to the great conjunction in 2015, no STEREO-B data since its contact was lost in October 2014, and occasional data gaps for various reasons. These events are excluded in the following analysis.

\item Multiple (22 events): No link can be established between the CME in question and the SEP event because of multiple CMEs and ESP events that occurred in succession. These events are excluded in the following analysis.

\end{itemize}

We note that the required threshold $v_{CME} \geq$ 900~km~s$^{-1}$ excluded some large SEP events. Four $>$10~MeV proton events with peak fluxes exceeding 10~pfu are not included in this study. The starting dates of these events are 2012 July 12, 2013 April 11, 2014 November 1, and 2015 October 29. The speeds of their associated CMEs were 885~km~s$^{-1}$, 861~km~s$^{-1}$, 740~km~s$^{-1}$, and 530~km~s$^{-1}$, respectively.  We also point out that our selected events are different from those in previous statistical studies of SEP events that were not restricted in CME parameters. For example, the fraction of CMEs with $v \geq$ 900~km~s$^{-1}$ was only 147/217 and 85/214, respectively, in the works by \citet{kahler_2013_apj} and by \citet{richardson_2014_sp}. Accordingly, these authors included SEP events as observed by Wind, SOHO and STEREO that were too weak to be observed by GOES; 
since the GOES background is higher than those of the other three missions.


\begin{figure*}
\gridline{\fig{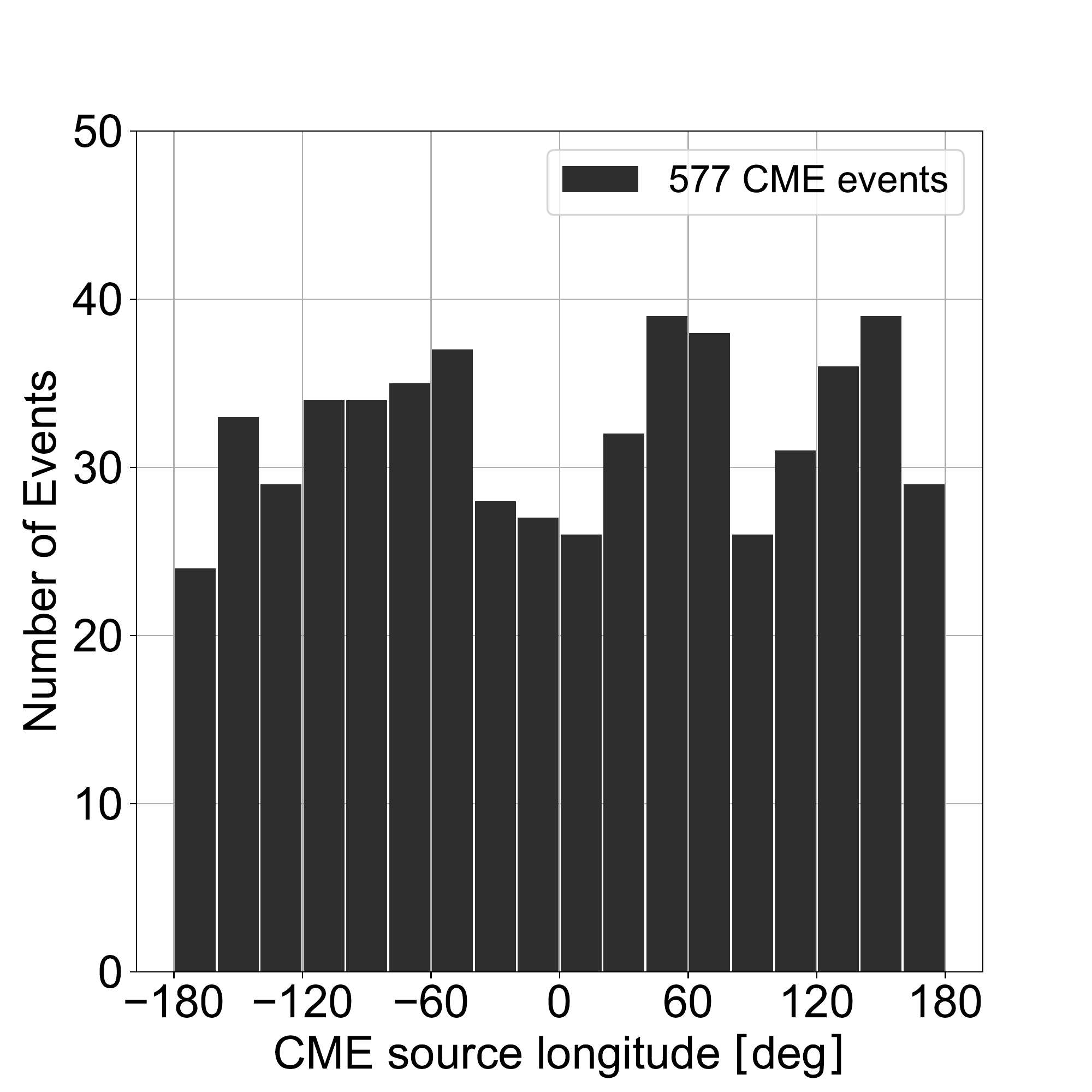}{0.33\textwidth}{(a)}
          \fig{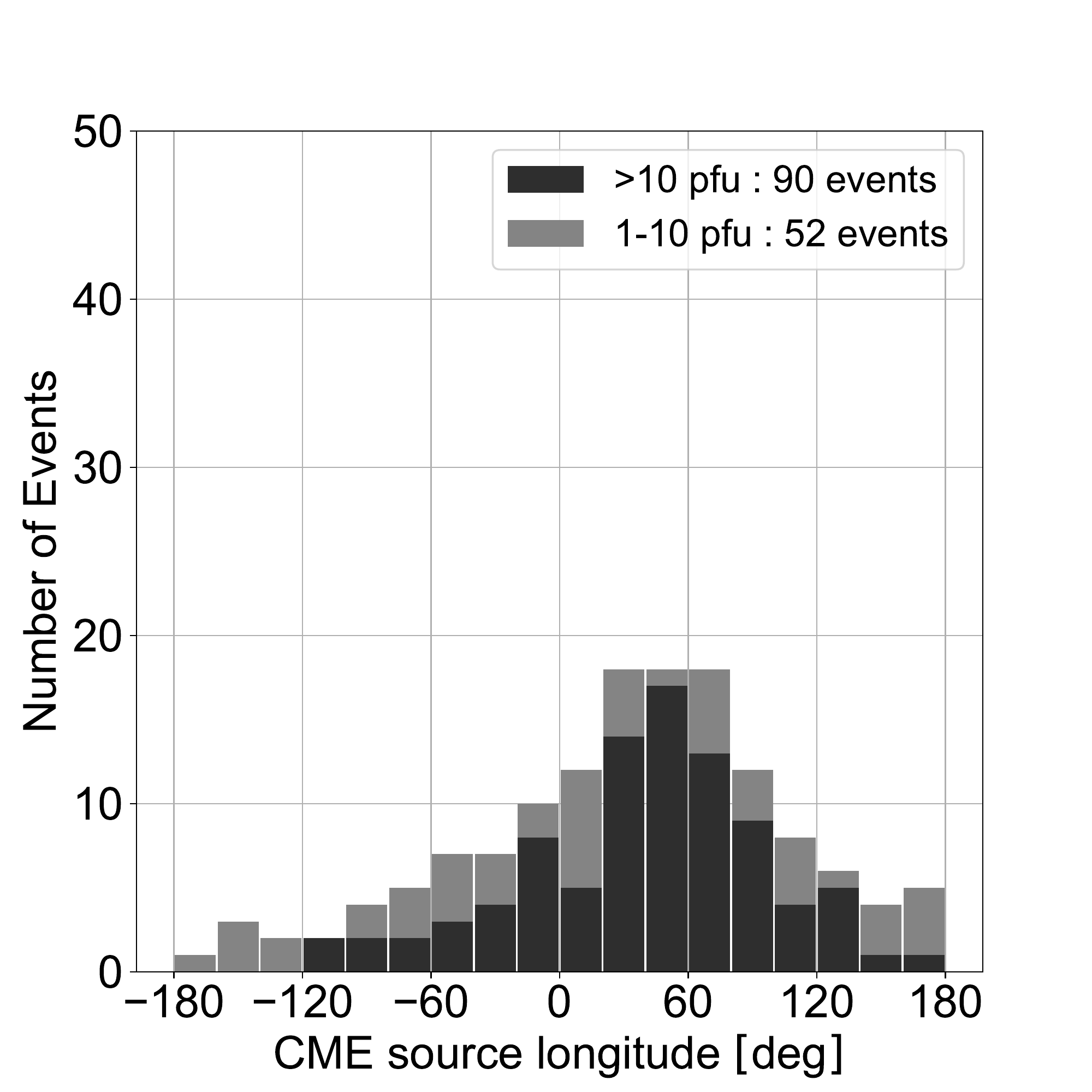}{0.33\textwidth}{(b)}
          \fig{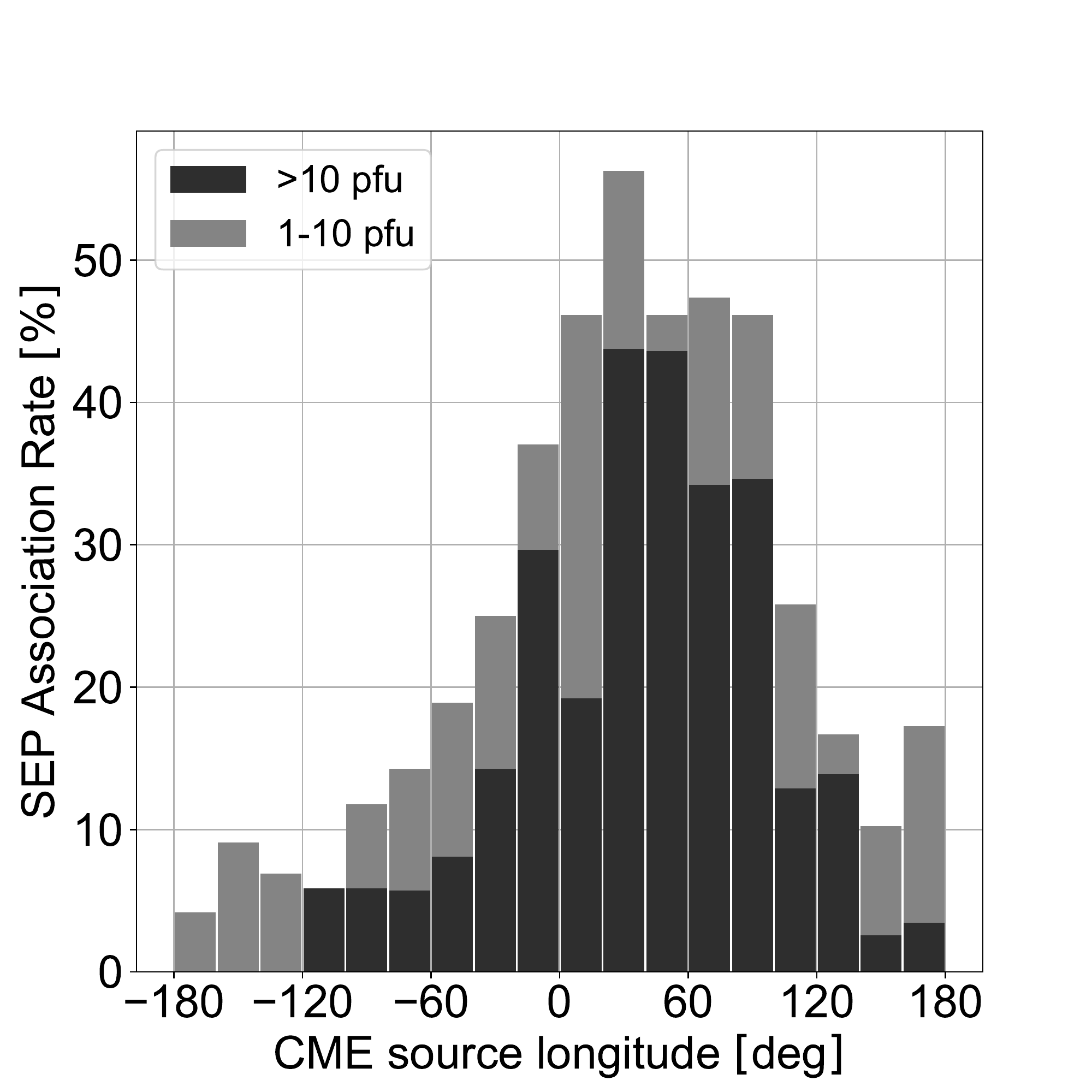}{0.33\textwidth}{(c)}}
\caption{Distribution of CMEs with source longitude (in 20$\arcdeg$ bins) relative to the observer.  (a) All the  CMEs, irrespective of their associations with SEP events. CMEs that occurred during high background periods are excluded (39 HiB$\_N$ events, with $N \geq$10, see Section 2). (b) Only the CMEs associated with a SEP event exceeding 10~pfu (black) and between 1 and 10~pfu (gray).  (c) The percentage of CMEs associated with a SEP event (the color usage same as (b)). Panels (b) and (c) exclude 10 and 2 HiB\_$N$ events (with $N \geq$10 and $N \geq$1) for $>$10~pfu and 1\,--\,10 pfu, respectively.
}
\label{ov_sl}
\end{figure*}

\section{SEP Association of Fast and Wide CMEs}
\begin{figure*}
\gridline{\fig{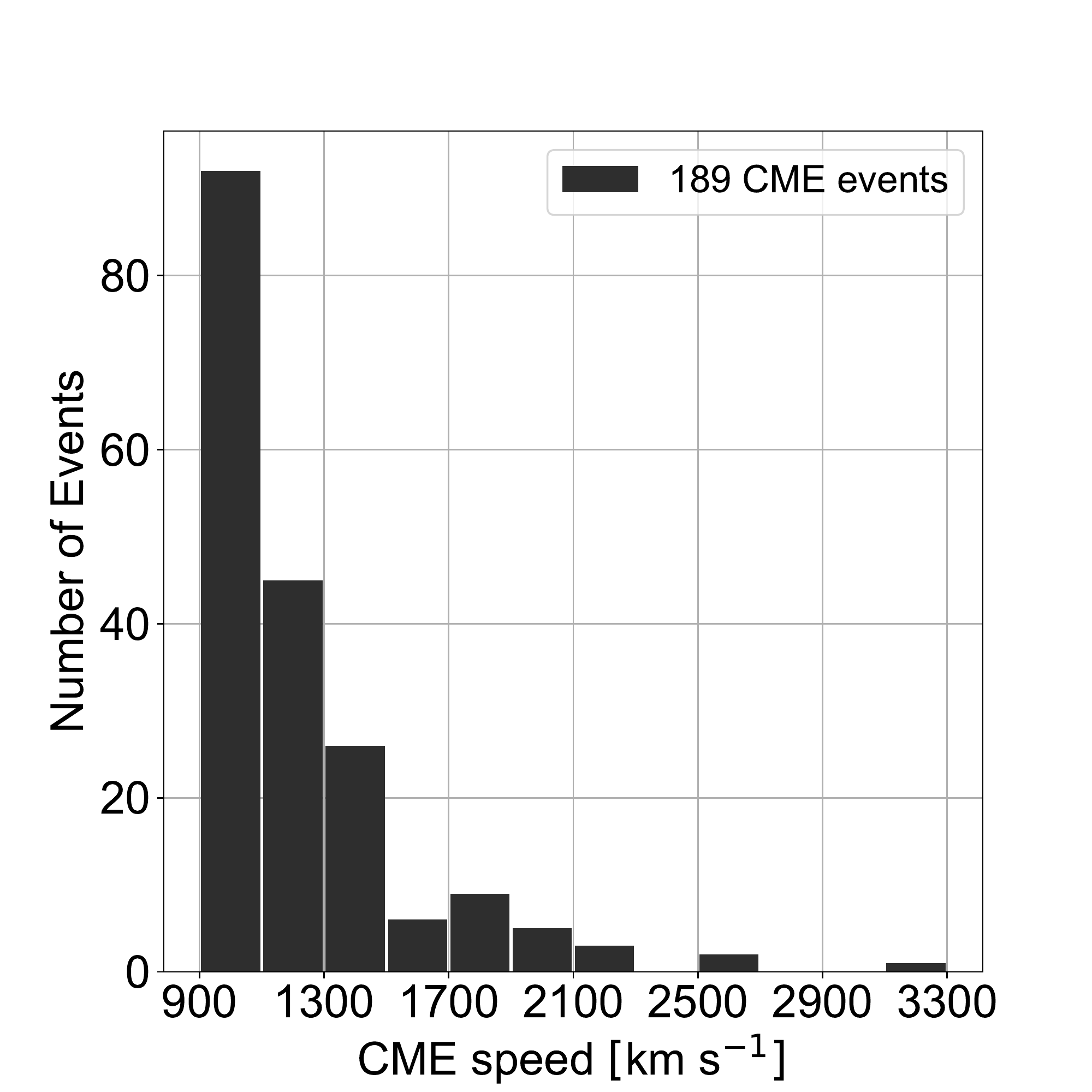}{0.33\textwidth}{(a)}
          \fig{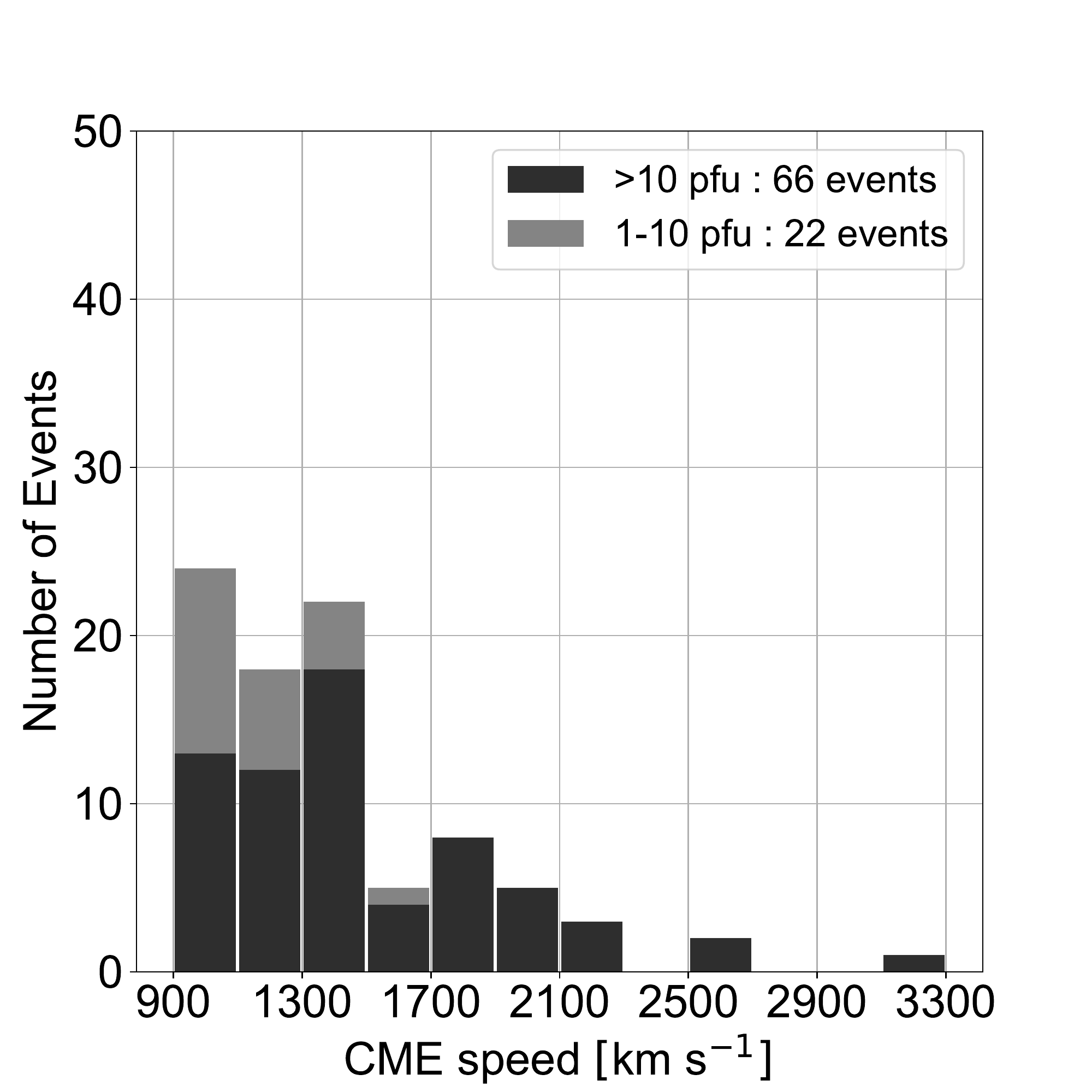}{0.33\textwidth}{(b)}
          \fig{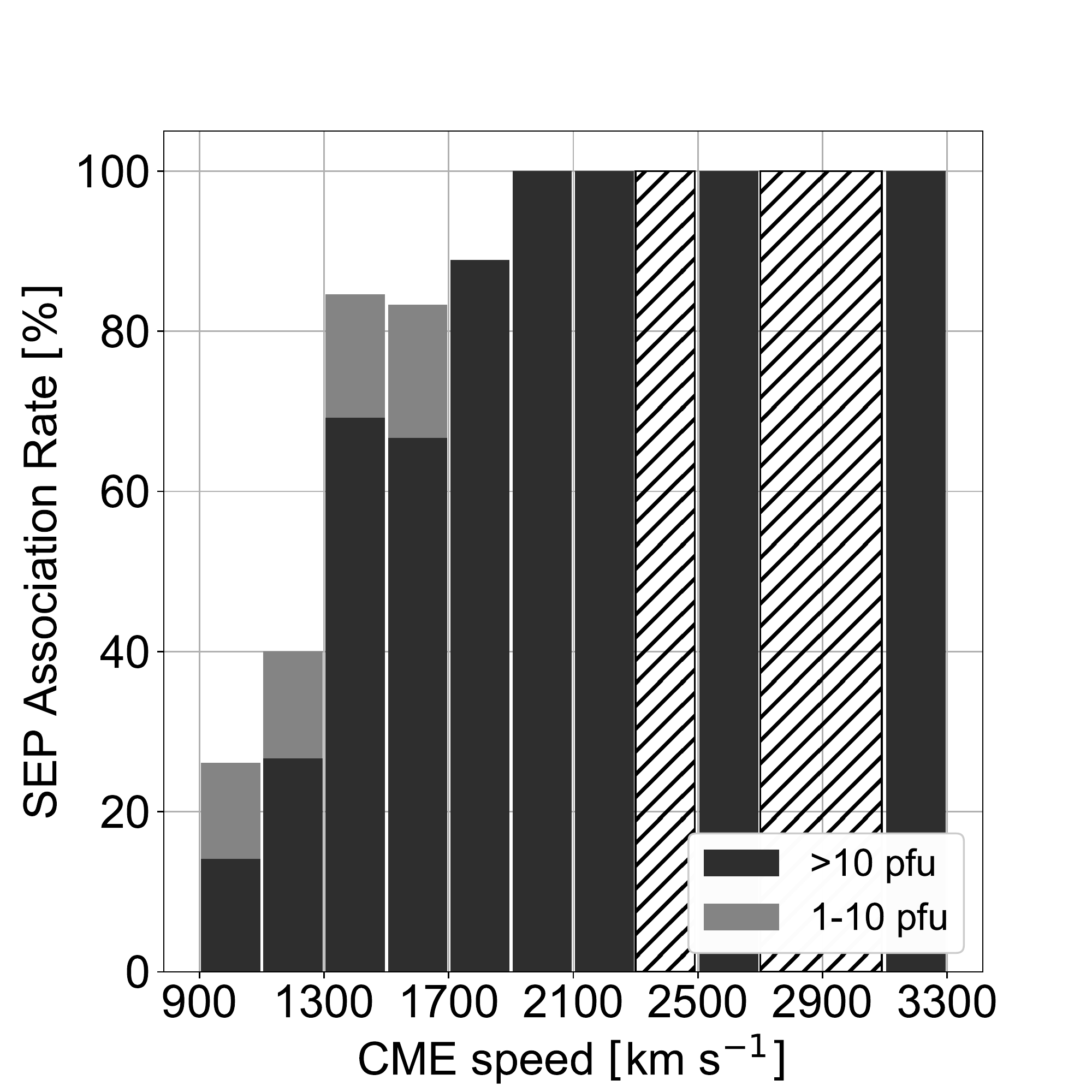}{0.33\textwidth}{(c)}}
\caption{Distribution of CMEs with speed in 200~km~s$^{-1}$ bins. Here we limit the CMEs to those from the source longitudes E20\,--\,W100. (a) All the  CMEs, irrespective of their associations with SEP events.(b) Only CMEs associated with a SEP event exceeding 10~pfu (black) 
and between 1 and 10~pfu (gray). (c) The percentage of CMEs associated with a SEP event (the color usage is the same as in (b)). 
Note that there were no CMEs between 2300~km~s$^{-1}$ and 2500~km~s$^{-1}$ and between 2700~km~s$^{-1}$ and 3100~km~s$^{-1}$, 
as indicated by the hatched areas in panel (c). As in Figure~{\ref{ov_sl}}, HiB$\_N$ events are excluded ($N \geq$10 in (a), $N \geq$10 and $N \geq$1 in (b) and (c), for $>$10~pfu and 1\,--\,10 pfu, respectively.
}
\label{ov_sp}
\end{figure*}

This work involves careful analyses of EUV images to locate the region from which each of the selected CMEs originated, using known low coronal signatures of CMEs \citep[e.g.,][]{zhang_2007_jgr, hudson_2011_jgr, nitta_2014_sp}.  
Figure~{\ref{ov_sl}} shows the distribution of our CMEs with the longitude of the source region relative to the observer. Note that the same CME can appear up to three times at different longitudes, as seen from Earth, STEREO-A, and STEREO-B.
In Figure~{\ref{ov_sl}}(a), we include all the CMEs, irrespective of their associations with SEP events,
except for 39 ``HiB$\_N$'' ($N \geq$10) events (Section 2). In these high background events, the association with a SEP event exceeding 10~pfu is uncertain, since it is possible that such an event could be buried under the background. Thanks to multi-spacecraft observations, we have at least 23 CMEs in each of the 20$\arcdeg$ bins.
Figure~{\ref{ov_sl}}(b) shows the distribution of SEP-associated CMEs, and Figure~{\ref{ov_sl}}(c) the ratio of SEP-associated CMEs to all CMEs.
In Figures~{\ref{ov_sl}}(b) and {\ref{ov_sl}}(c), we show the association of CMEs with SEP events as defined by two thresholds for the peak flux: 10~pfu and 1~pfu.  
Figure~{\ref{ov_sl}}(c) shows that the SEP association rate is elevated in the range of longitude of E20\,--\,W100.  We therefore refer to this range of longitude as ``well-connected'' in this section (i.e., Figures~{\ref{ov_sp}} and {\ref{other_cor}}(a)). This is broader than what was shown in previous studies on the longitudinal distributions of SEP events \citep[e.g.,][]{smart_1996_asr, laurenza_2009_sw}.  It is possible that the broader distributions in Figure~{\ref{ov_sl}}(c) may be characteristic of our SEP events measured at $>$10~MeV, most of which come from solar cycle 24.


In Figure~{\ref{ov_sp}}, we show the distribution of CMEs with speed in 200~km~s$^{-1}$ bins. Here we limit the CMEs to those from the well-connected longitudes (E20\,--\,W100).   
As expected, there are more CMEs that are slower (but $\geq$ 900~km~s$^{-1}$), although
the SEP association rate rises sharply with the CME speed. 
All the CMEs faster than 2100~km~s$^{-1}$ are associated with a SEP event.

\begin{figure*}
\gridline{\fig{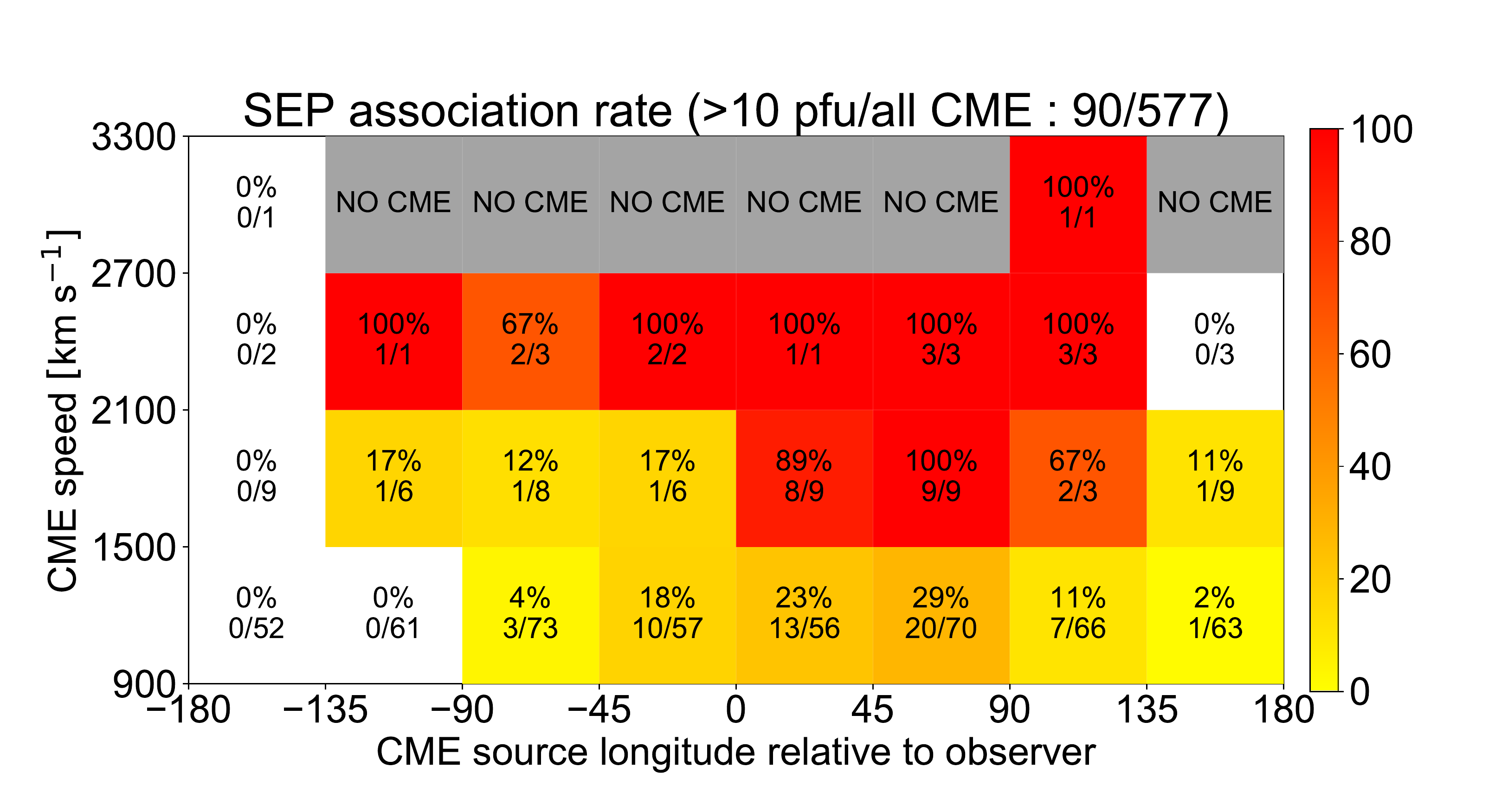}{0.5\textwidth}{(a)}
          \fig{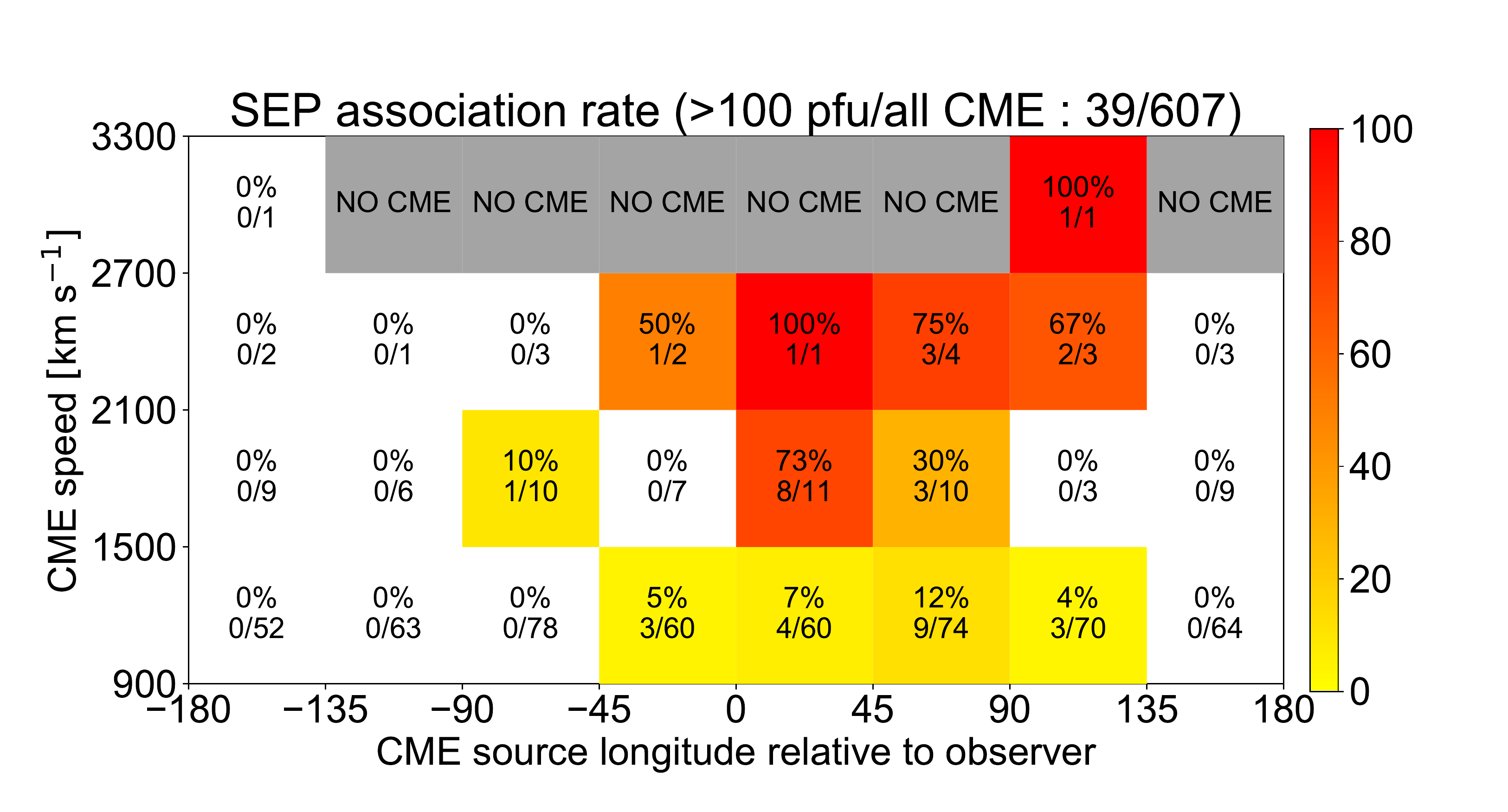}{0.5\textwidth}{(b)}}
\caption{SEP heat maps, showing how the SEP association rate varies with the source longitude and speed of the CME. (a) and (b) show, respectively, a heat map for SEP events $>$10~pfu and $>$100~pfu. Higher (lower) association rates are shown in red (yellow). Cells in gray indicate no CMEs in those ranges of speed and source longitude.  Cells in white indicate no SEP association.
Different numbers of the HiB\_$N$ events are excluded in (a) and (b), depending on $N$; $N \geq$10 and $N \geq$100 for (a) and (b), respectively (see text).
}
\label{hz}
\end{figure*}

The dependence of the SEP association rate of CMEs on both the longitude of the source region and the CME speed can be visualized in heat maps, as shown in Figure~{\ref{hz}}. Here the maps come from a 2D array of the longitude relative to the observer (in 45$\arcdeg$ bins) and the speed (in 600~km~s$^{-1}$ bins) of the CME. Colors from yellow to red represent low to high association rates, as indicated in the color bars in the figure. Note that the SEP association rates have large uncertainties except for the bottom row because of the limited number of events. Figures~{\ref{hz}}(a) and {\ref{hz}}(b) are heat maps for the SEP association rate for SEP events with threshold peak fluxes $>$10~pfu and $>$100~pfu, respectively. The number of all the CMEs and of the SEP-associated CMEs, together with the association rate, is also indicated in each cell.  
As in the previous figures, we exclude the HiB\_$N$ events that prevent us from determining the association of the CME with a SEP event.  For the association with SEP events exceeding 10~pfu (panel (a)), 39 events with $N \geq$10 are excluded.  For SEP event exceeding 100~pfu (panel (b)), 9 events with $N \geq$100 are excluded.  As a result, the number of all the CMEs is greater in panel (b). 
Note that there was only one CME faster than 2700~km~s$^{-1}$, which occurred on 2017 September 10. Its speed was 3163~km~s$^{-1}$. This CME appears twice in each heat map, since it occurred after the contact with STEREO-B was lost. Its source location was S09W90 from Earth.  
This translates to E142 from STEREO-A, at which only a weak SEP event was observed that seemed to be directly linked to the CME.  The $>$10~MeV proton flux peaked on September~12\footnote{A higher ($>$100~pfu) $>$10~MeV proton flux was seen on September~14, but this appears to have been due to a stream interaction region rather than to the CME on September~10 \citep[see][]{guo_2018_sw}.}, and the peak flux was less than 10~pfu.  Therefore  the top left cell indicates no SEPs in Figures~{\ref{hz}}(a) and {\ref{hz}}(b). 
There were fewer CMEs with higher speeds, and based on limited statistics, it appears that faster CMEs are associated with SEPs over wider longitudes.  For the association rates of CMEs with SEP events for which the peak exceeded 10~pfu, we can confirm that faster CMEs contribute more to the SEP association rate.  For example,  Figure~{\ref{ov_sl}}(c) shows that the SEP association rate of CMEs around W60 is $\sim$40\%, which results from the 100\% association rate for CMEs faster than 1500~km~s$^{-1}$ offsetting the lower association rate for slower CMEs.  Figure~{\ref{hz}}(b) shows that the CMEs associated with $>$100~pfu SEP events not only are much rarer but also are limited to higher speeds and narrower ranges of longitude.

\begin{figure*}
\gridline{\fig{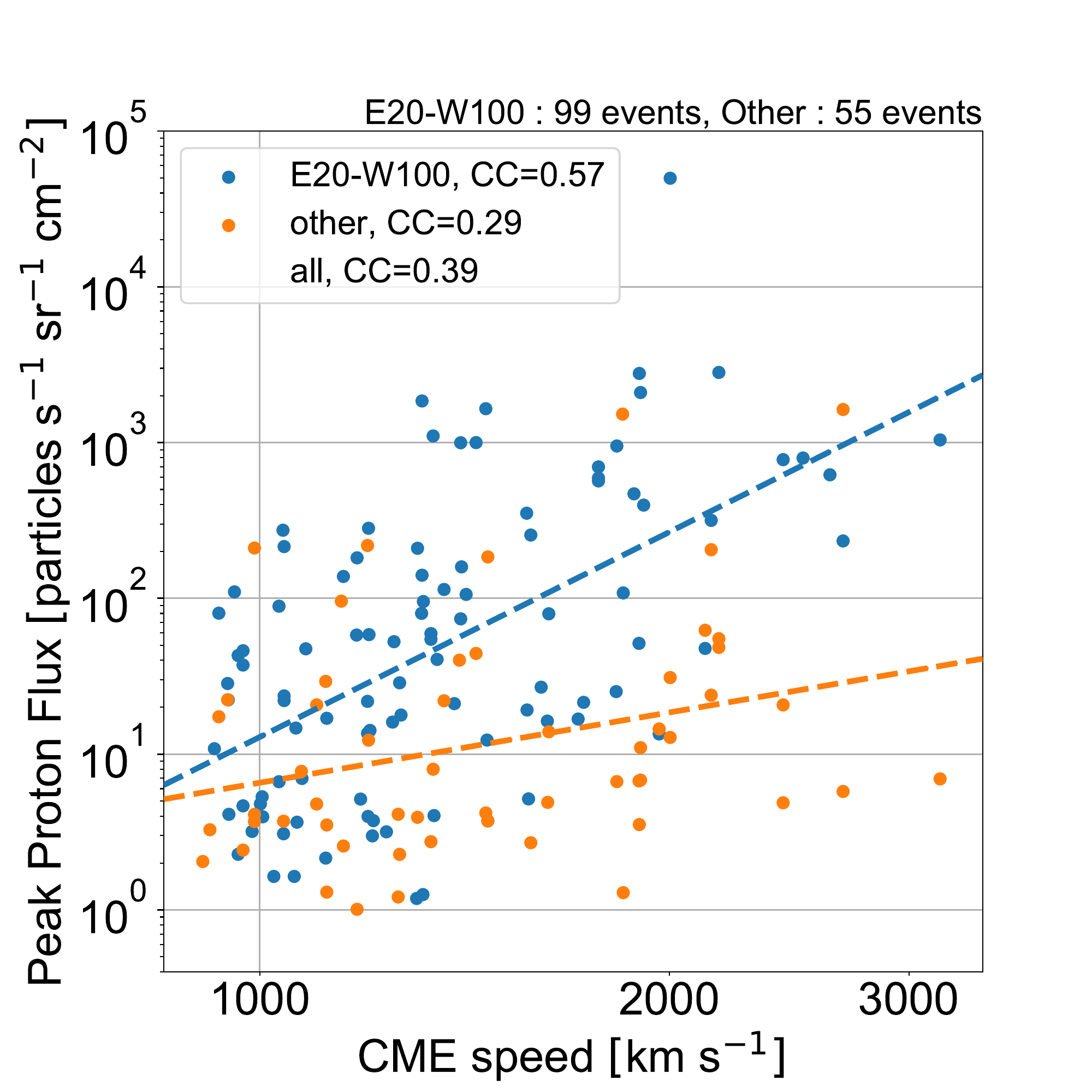}{0.33\textwidth}{(a)}
          \fig{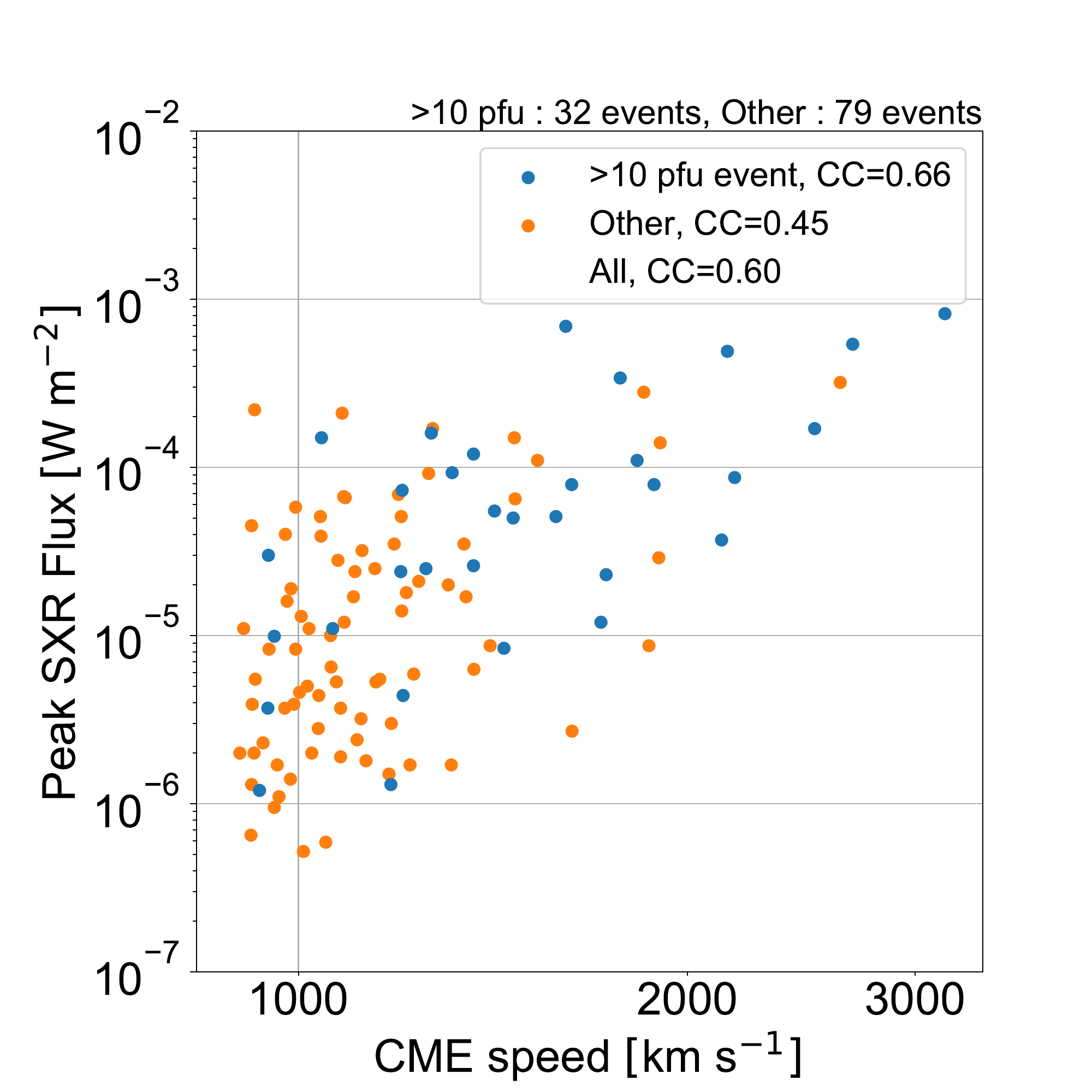}{0.33\textwidth}{(b)}
          \fig{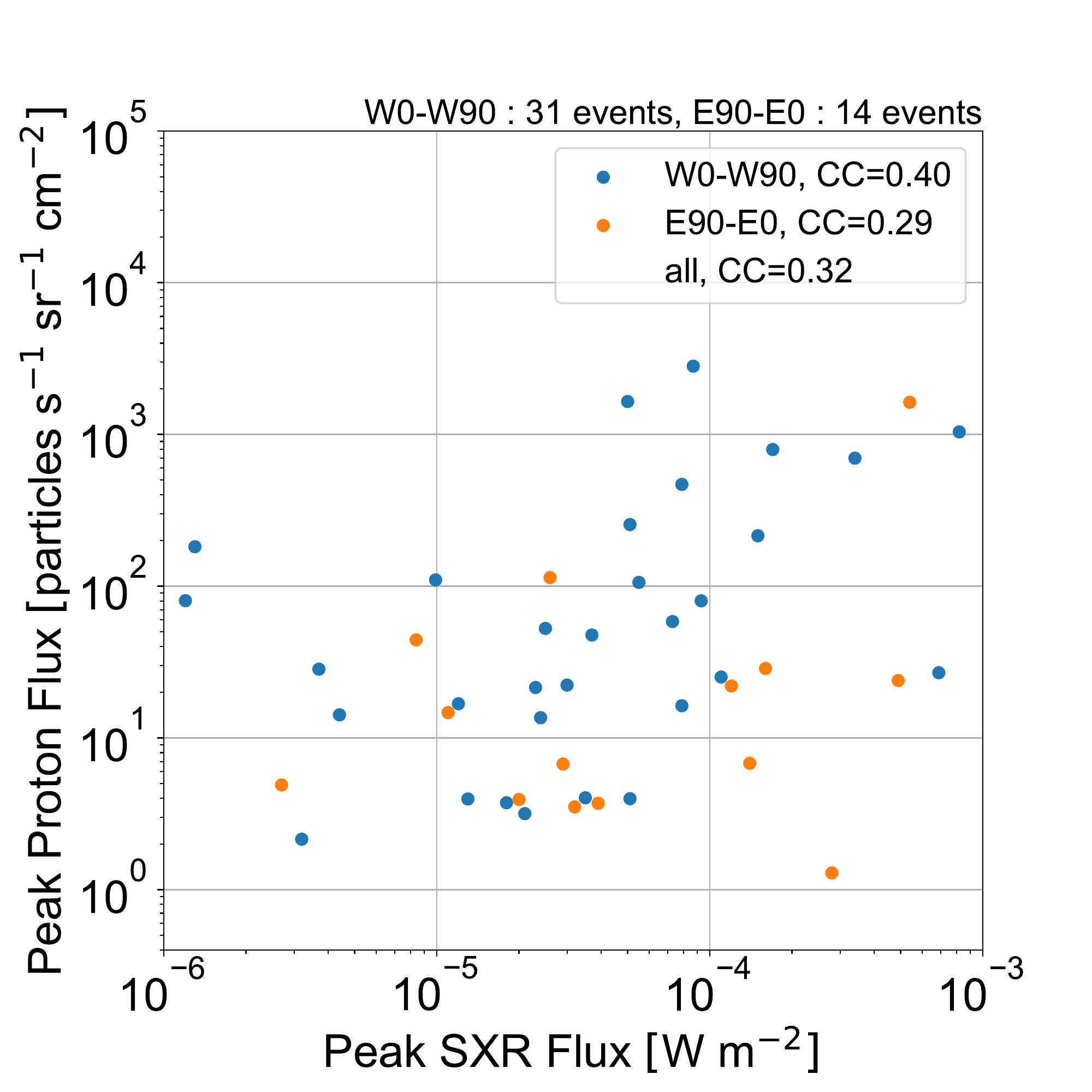}{0.33\textwidth}{(c)}}
\caption{(a) Correlation between CME speed and peak proton flux. Each symbol
         shows whether or not the events come from well-connected longitudes.
         The dashed lines are the regression lines for each group.
         All the events in which the peak proton flux was measured are included (``Good'', ``Contaminated'', and ``HiB\_$N$'').
         (b) Soft X-ray (SXR) flux vs CME speed as observed by GOES.
         This figure contains only the data from frontside events at longitudes E90-W90
         from GOES.
         The symbols show whether or not each CME is associated with a $>$10~pfu SEP event.
         (c) Peak proton flux vs soft X-ray flux (both from GOES) for frontside events.
         }
\label{other_cor}
\end{figure*}

In space weather applications, the SEP peak flux is routinely forecast using basic properties of CMEs and flares. In Figure~{\ref{other_cor}}, we show the relations among the SEP peak flux, the CME speed, and the magnitude of the associated flare.  
Figure~{\ref{other_cor}}(a) shows the relation between the peak proton flux and the CME speed.
Even though they appear to be weakly correlated, especially for those events in the well-connected longitudes (plotted in blue), there is considerable scatter, as found in past studies \citep[e.g.,][]{kahler_2001_jgr}; the peak proton flux for the same CME speed can vary by three orders of magnitude, even though we limit the CMEs to fast ones ($v_{CME}~\geq$~900~km~s$^{-1}$) in this study.
This is unlike \citet{kahler_2001_jgr}, who included even CMEs slower than 200~km~s$^{-1}$.  
The regression lines of each group of longitudes are also shown in Figure~{\ref{other_cor}}(a). The peak proton flux tends to be higher for the SEP events from well-connected longitudes. The difference is about one order of magnitude for a CME with 2000~km~s$^{-1}$.

In Figure~{\ref{other_cor}}(b), we plot the peak soft X-ray (SXR) flux of the associated flare vs the CME speed, considering separately those CMEs associated with, and not associated with, $>$10~pfu protons.  There is a somewhat higher correlation for SEP-associated CMEs, which tend to be faster (Figure~{\ref{ov_sp}}(a)).  This relation may largely reflect the big-flare syndrome \citep{kahler_1982_jgr}.
Figure {\ref{other_cor}}(c) shows a weak correlation between the peak SXR flux and peak proton flux,
irrespective of whether the source region is in the western or eastern hemisphere. Note that this plot contains only flares that were associated with fast and wide CMEs. The scatter would be much more pronounced if all flares were included irrespective of their associations with CMEs.

\section{SEP Timescales} \label{sec:ts}
The timescales are also important properties of SEP events.  Typical questions include: When does the SEP event start? How fast does it rise to the peak? And how long does it stay at a high level?
We measured the key times of SEP events only when the onset was clearly found.  They are labeled either ``Good'' or ``Contaminated'' in Table~{\ref{allcme}}, not including events that have high pre-event background. If the quality in Table~{\ref{allcme}} is ``Good''--- we measured the following four times in this study:
the SEP start time, the SEP half-peak (start), the SEP peak time,
and the SEP half-peak (end). We measured all times on a log scale plot.
We defined the SEP start time manually as the time when the proton
flux increased above the background, typically in three consecutive 5~min intervals.
The SEP peak time is usually the time after the SEP start time when the proton flux is the
highest. More complex cases are described below.
The SEP half-peak (start) and SEP half-peak (end), respectively, were automatically
extracted as the times when the proton flux first exceeded---and last went below---half of the flux at the SEP peak time.  

Many SEP events show simple rise-and-fall time profiles, and we were able to measure the four times defined above unambiguously for these events.  However, in some events the proton flux time profile exhibited a second or even third peak, following a plateau or plateaus after the first peak. In such events, we usually took the time of the first peak as the
SEP peak time. This is because later peaks may be produced by transport effects and thus may not directly reflect the CME properties.  Other SEP events show a clear onset, but subsequent time profiles are contaminated either by another SEP event due to a later CME or by an ESP event locally produced by the passage of the shock wave driven by the present or an earlier CME.  They are labeled ``Contaminated'' in Table~{\ref{allcme}}. We did not measure the SEP half-peak (end) for any of these 26 events, but in all of them the SEP peak was clearly seen, allowing us to measure the SEP start time, SEP half-peak (start) and SEP peak time.

\begin{figure*}
\gridline{\fig{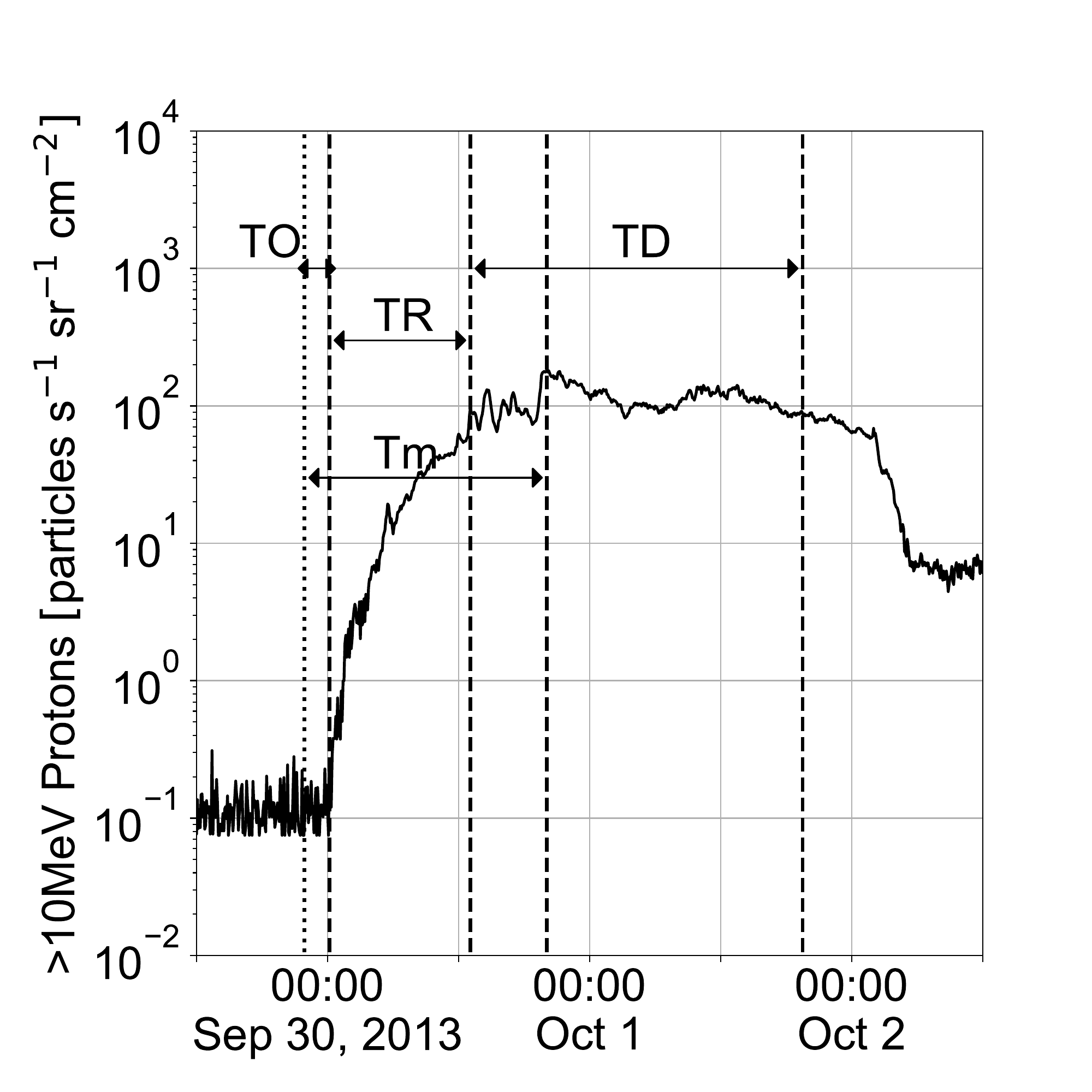}{0.5\textwidth}{}}
\caption{An example showing the four timescales defined in the text.  The dashed lines, from left to right, are the SEP start time, the SEP half-peak time (start),
the SEP peak time, and the SEP half-peak time (end).
The dotted line is the CME launch time from SOHO LASCO CME catalog.}
\label{exts}
\end{figure*}

From these times, we calculated the four timescales listed in Table~{\ref{allsep}}.
They are illustrated in Figure~\ref{exts} using an example of a real SEP event.
Here, TO is the onset time, which measures how quickly the SEP event starts at 1~AU (defined as the SEP start time) after the CME launch at 1~R$_{\odot}$;
TR is the rise time from the SEP start time to
the SEP half-peak (start); 
Tm is the full rise time from the SEP start time to the SEP peak time; and 
TD is the duration, i.e., the length of time during which the proton flux stays above half the peak value,
which is between SEP half-peak (start) and SEP half-peak (end).
These notations---TO, TR, and TD---were adopted by \citet{kahler_2013_apj}.
He also used OR, which is the sum of TO and TR; that is, 
the time from the CME launch at 1~R$_{\odot}$
to SEP half peak (start).
The timescale Tm is a redefinition of $\Delta T_{m}$ that was first used by \citet{vanhol_1975_sp}. 

\begin{figure*}
\gridline{\fig{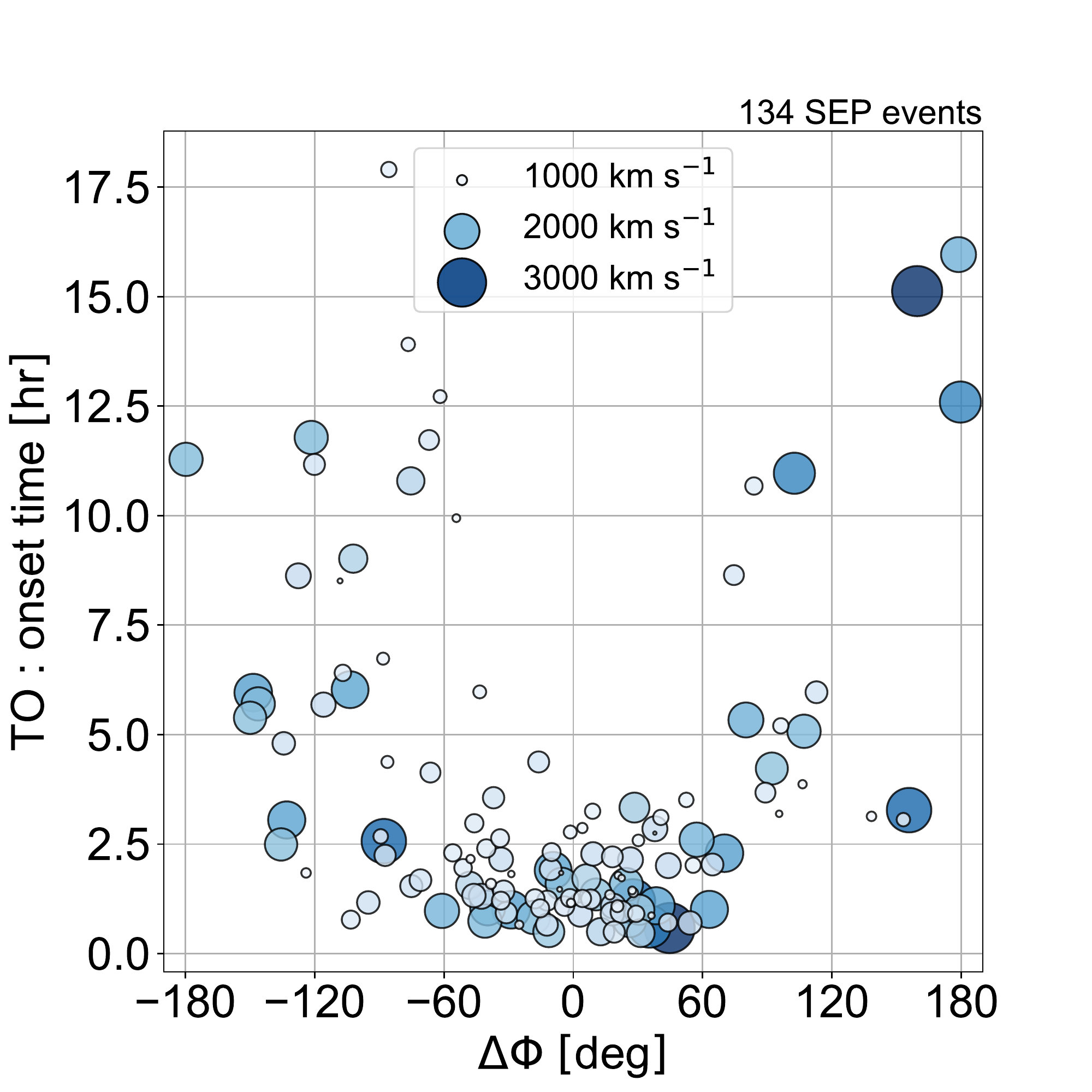}{0.33\textwidth}{(a)}
          \fig{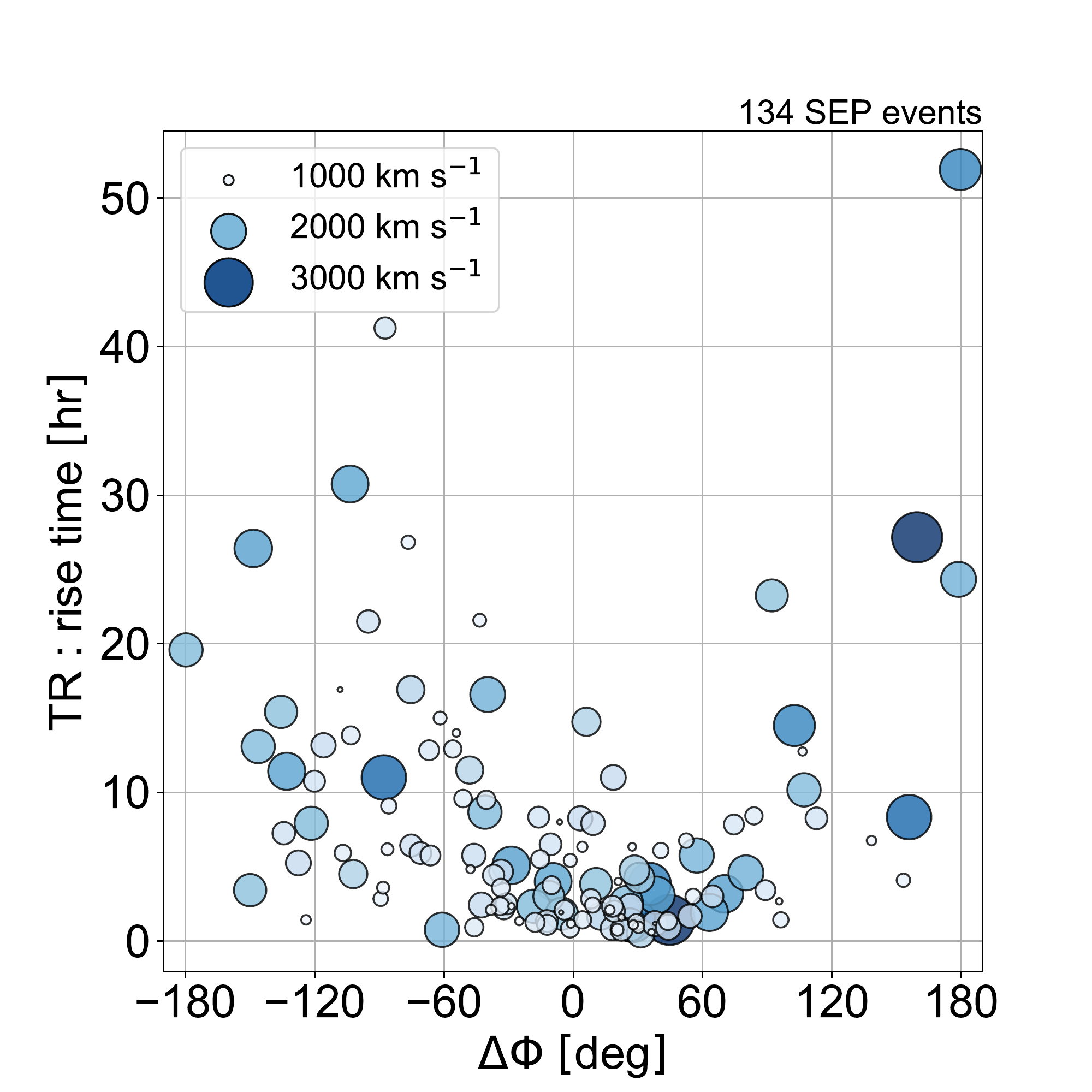}{0.33\textwidth}{(b)}
          \fig{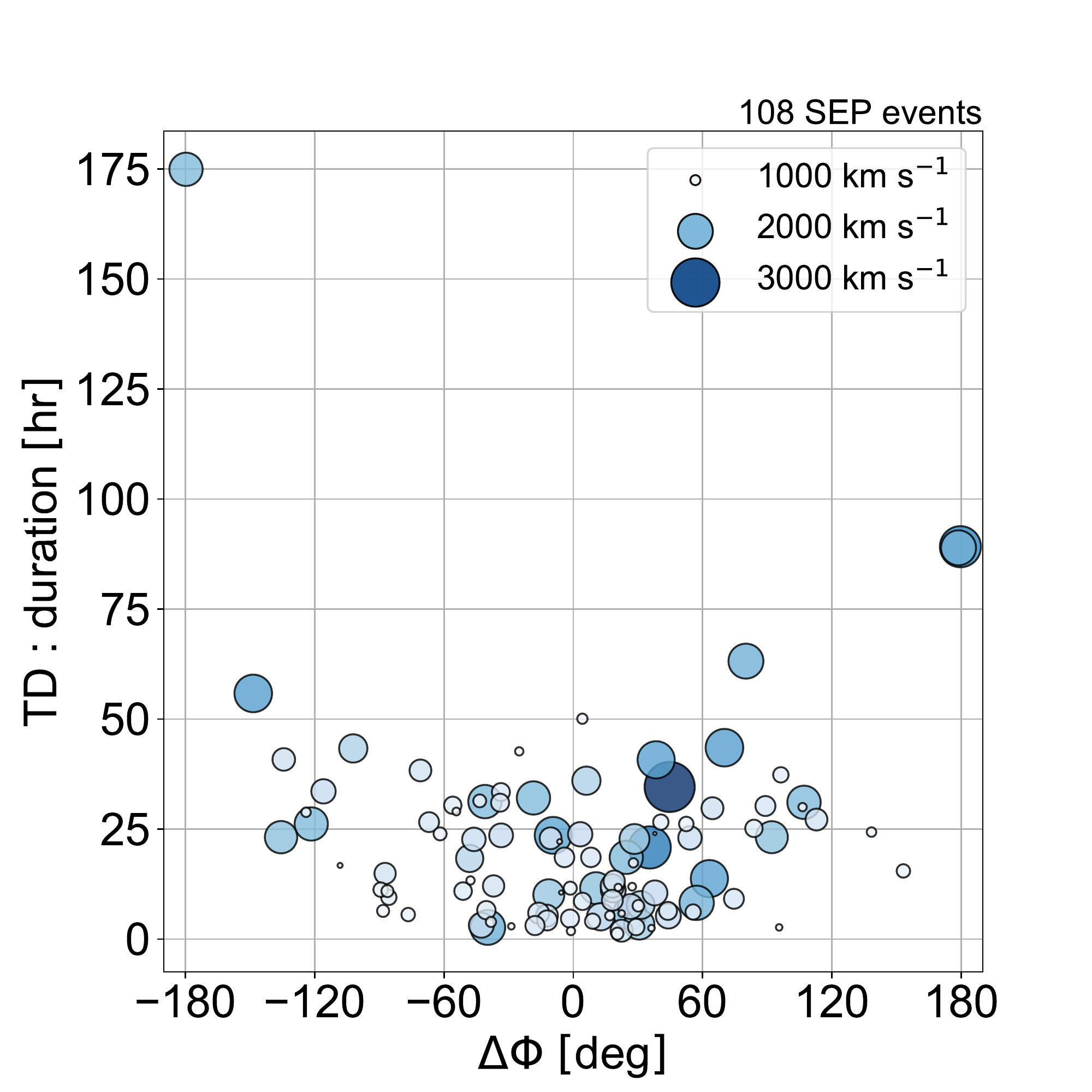}{0.33\textwidth}{(c)}
          }
\caption{(a) SEP onset time from the CME launch (TO) vs the longitude of the CME source region
relative to the footpoint of the Parker spiral.
(b) and (c) The same plot, but for the SEP rise time TR and SEP duration TD.
The size and darkness of the circles correspond to the CME speed.
Almost all the SEP events listed in Table~{\ref{allsep}} are included in panel (a) and (b),
except for the one on 2006 December 13 observed at STEREO-A and STEREO-B. This is because 
both STEREO were located near Earth and their data points are consistent with that of GOES.
Events in the Contaminated category are excluded in panel (c) because TD cannot be measured.
}
\label{ts_ps}
\end{figure*}

Here we study the correlation of the SEP timescales TO, TR and TD with CME source longitude and speed.
The plots in Figure~{\ref{ts_ps}} show the correlation of TO and TR with the CME source longitude relative to the footpoint of the Parker spiral ($\Delta \Phi$).
We calculated the longitude on the solar surface of the nominal footpoint of the Parker spiral that was
connected to GOES, STEREO-A, and STEREO-B, using the solar wind speed around the time of the SEP onset, as sampled by Wind and STEREO.  When the Wind data were missing, we used data from the Advanced Compositional Explorer (ACE).
Some events had to be dropped because we were not able to calculate the longitude of the Parker spiral footpoint due to the unavailability of solar wind data.
These timescales are plotted vs $\Delta \Phi$ in Figures~{\ref{ts_ps}}(a\,--\,c), where larger and darker circles indicate faster CMEs.
Figure~{\ref{ts_ps}}(a) shows that CMEs from regions within 60$\arcdeg$ in longitude from the footpoint of
the Parker spiral tend to be associated with SEPs with short onset times. 
In this region of $\Delta \Phi$, TO is almost always shorter than five hours but longer than one hour. Note that it takes 1.15 hours for 10~MeV protons to travel the distance of 1.2~AU, which is often used as a typical path length for the Parker spiral corresponding to a solar wind speed of $\sim$400~km~s$^{-1}$.
The longest TO is about 18 hours for an event that is far outside this range of $\Delta \Phi$.
In Figure {\ref{ts_ps}}(b), we find a similar trend for TR with $\Delta \Phi$, but with more scatter, even for $| \Delta \Phi |~<$~60$\arcdeg$.  In Figure~{\ref{ts_ps}}(c), TD is characterized by a broad distribution for most ranges of $\Delta \Phi$, with occasional high values.

\begin{figure*}
\gridline{\fig{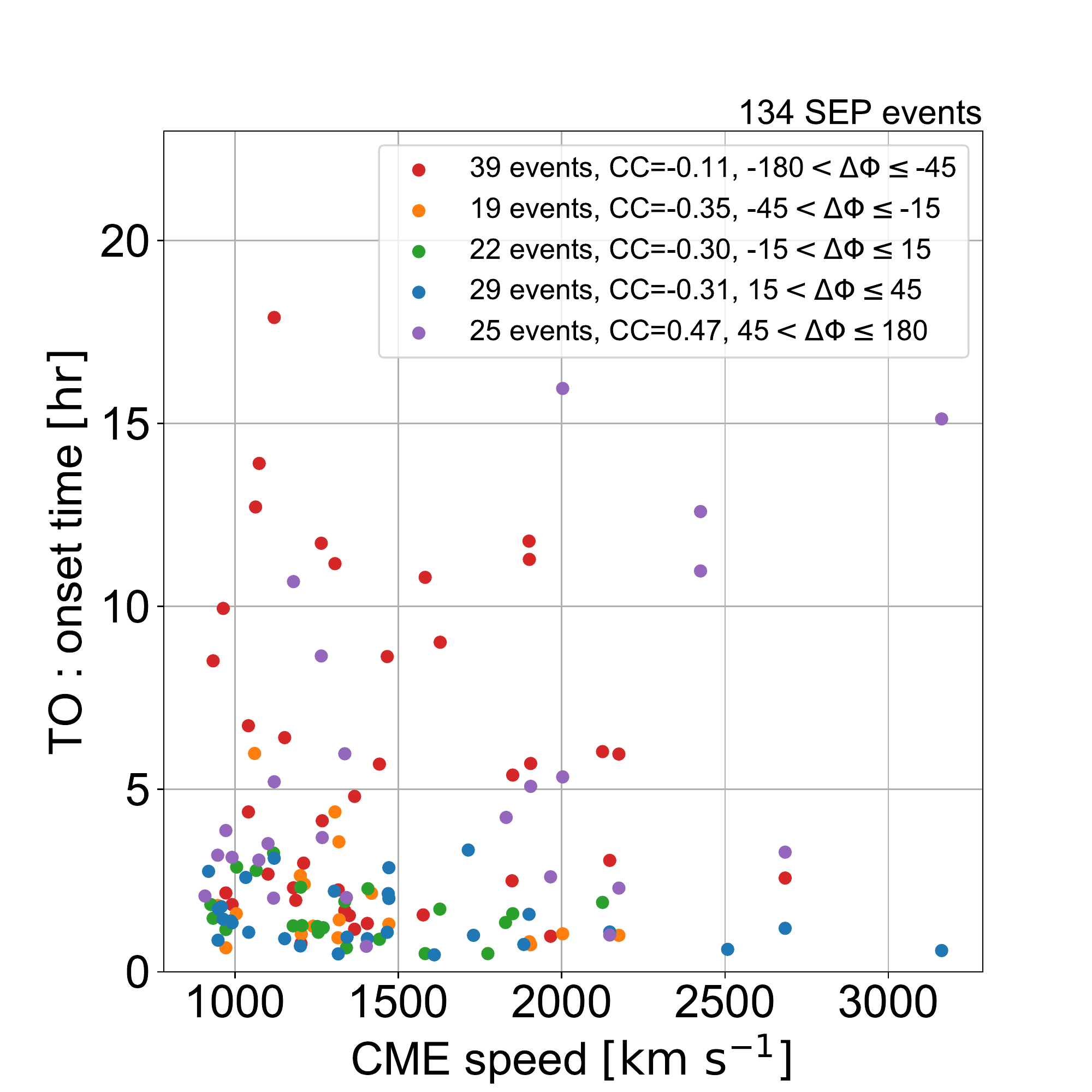}{0.5\textwidth}{(a)}
          \fig{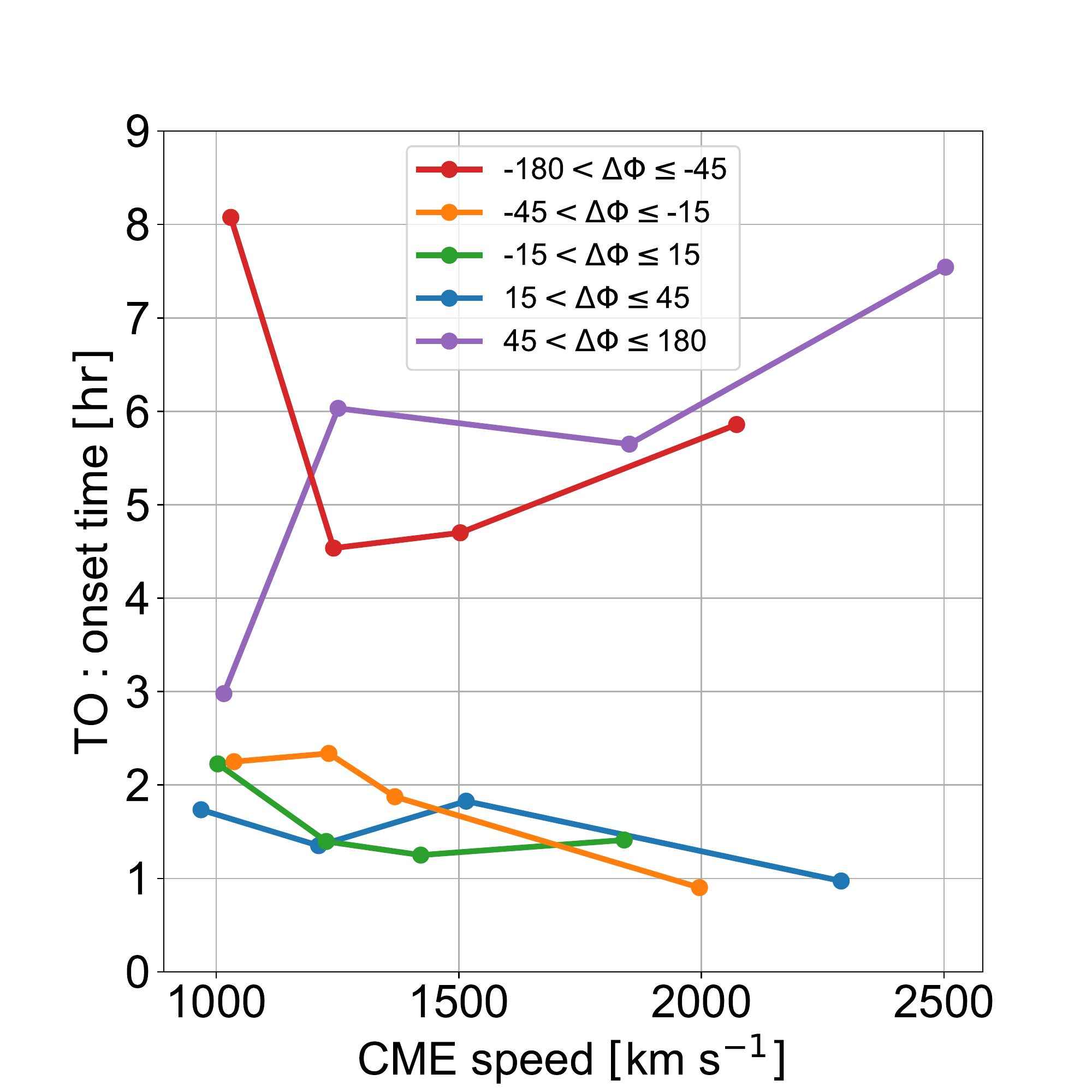}{0.5\textwidth}{(b)}}
\caption{SEP onset time TO vs CME speed ($v_{CME}$). Different colors are used for events in different ranges of longitude relative to the 
footpoint of the Parker spiral.
Individual TO values are plotted in the left panel.  In the right panel, averaged TO values are plotted after re-grouping all the data into four representative CME speeds. 
The number of data points is the same as in Figure {\ref{ts_ps}}(a).
}
\label{TO_speed}
\end{figure*}

\begin{figure*}
\gridline{\fig{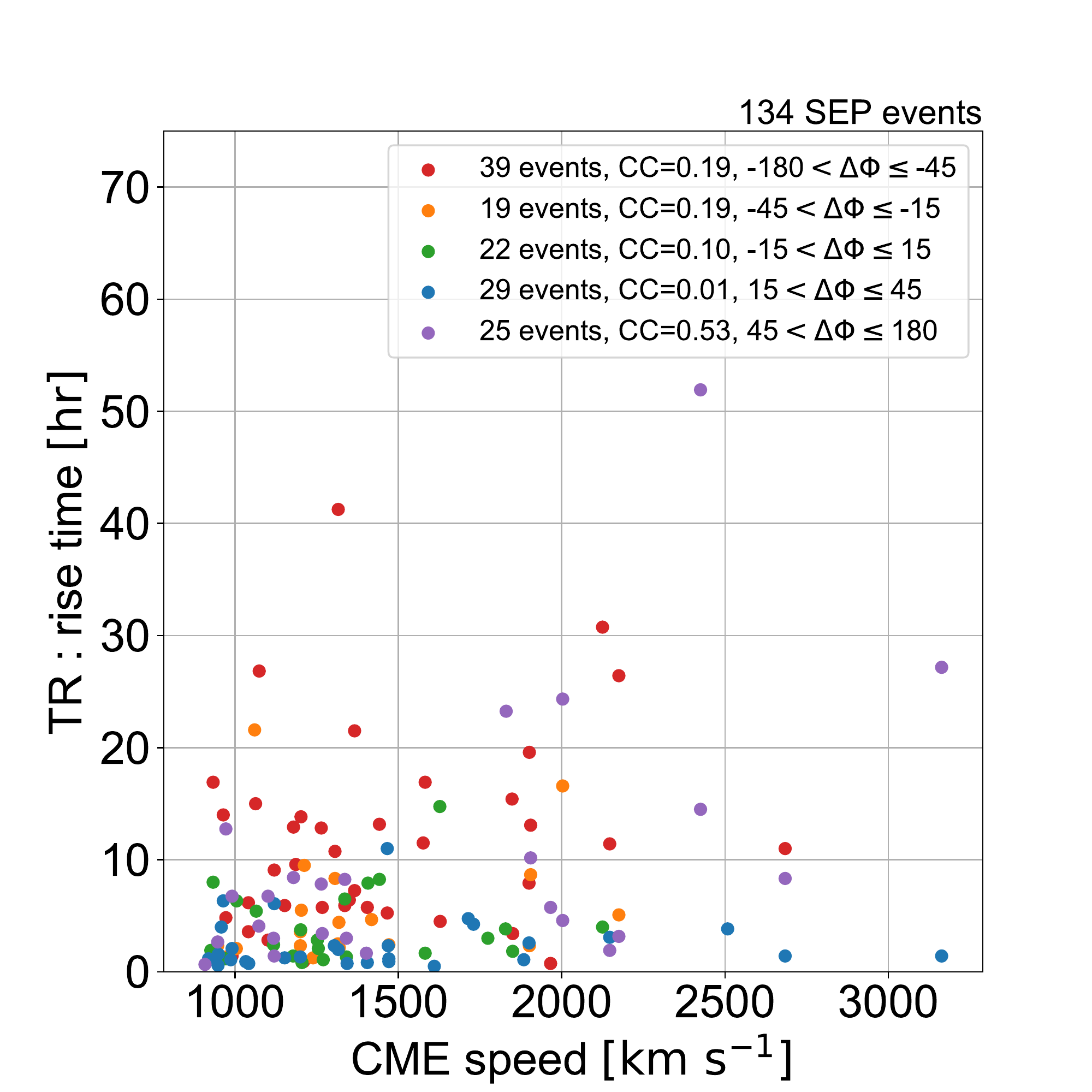}{0.5\textwidth}{(a)}
          \fig{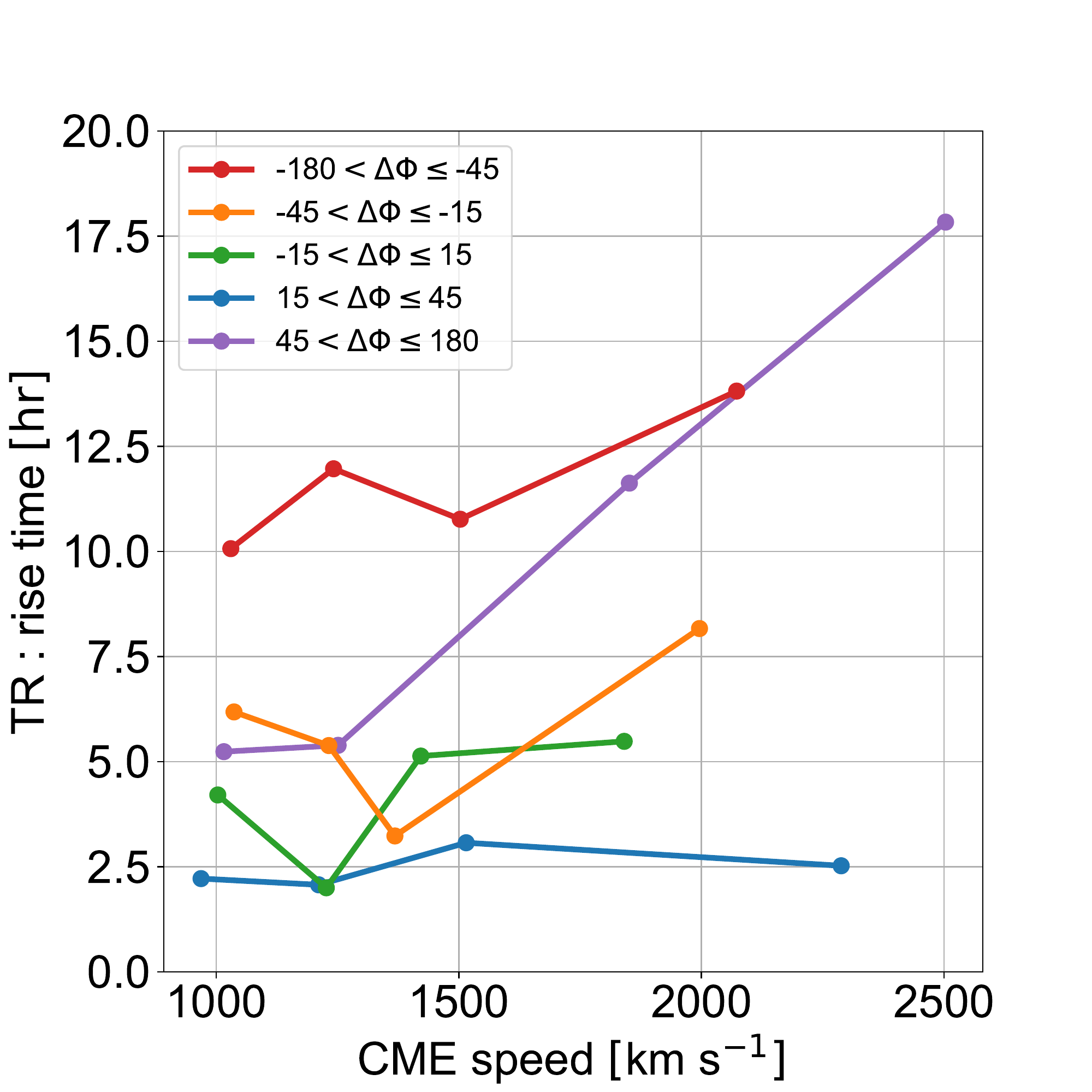}{0.5\textwidth}{(b)}}
\caption{Same as Figure {\ref{TO_speed}}, but for the rise time TR.
The number of data points is the same as in Figure {\ref{ts_ps}}(b).
}
\label{TR_speed}
\end{figure*}

\begin{figure*}
\gridline{\fig{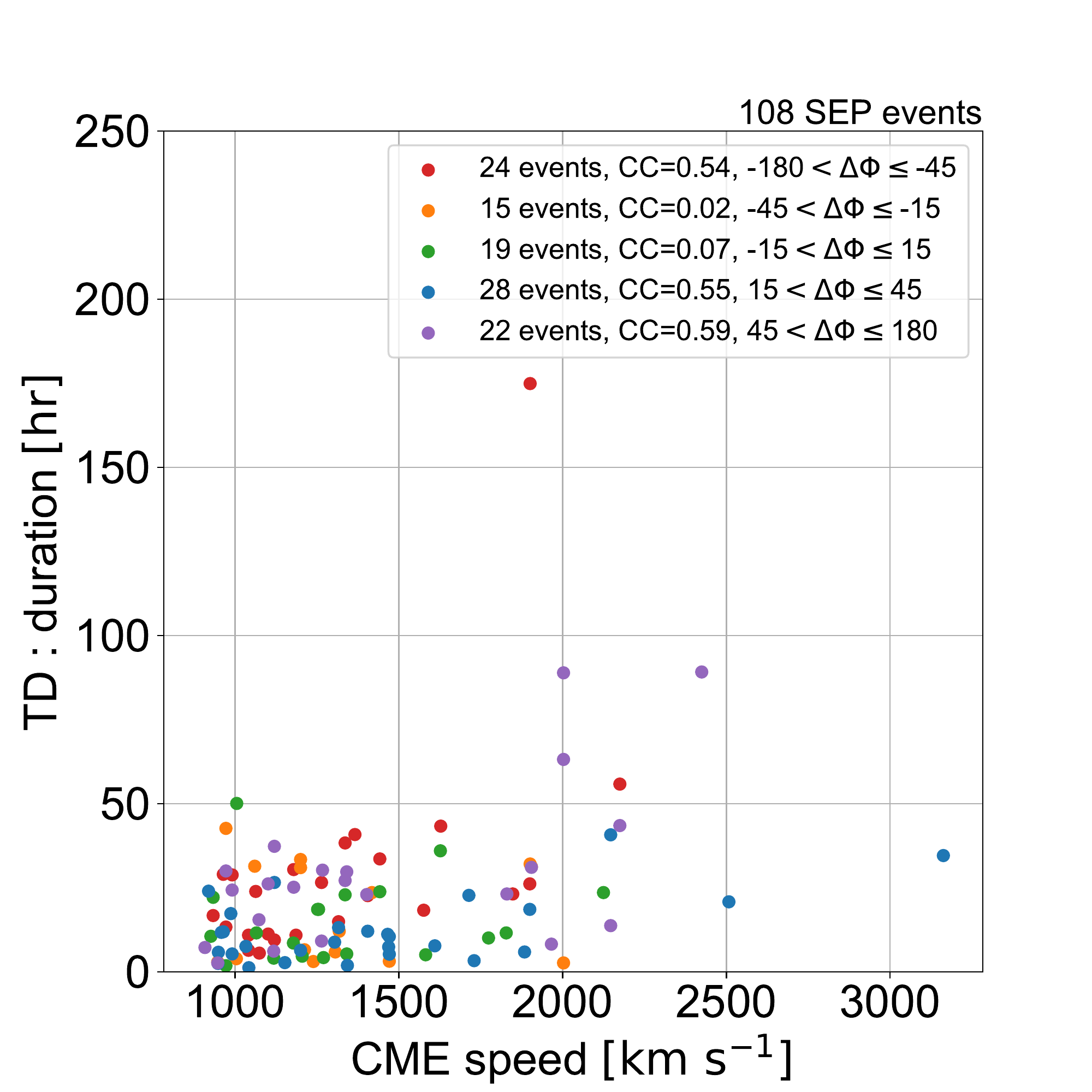}{0.5\textwidth}{(a)}
          \fig{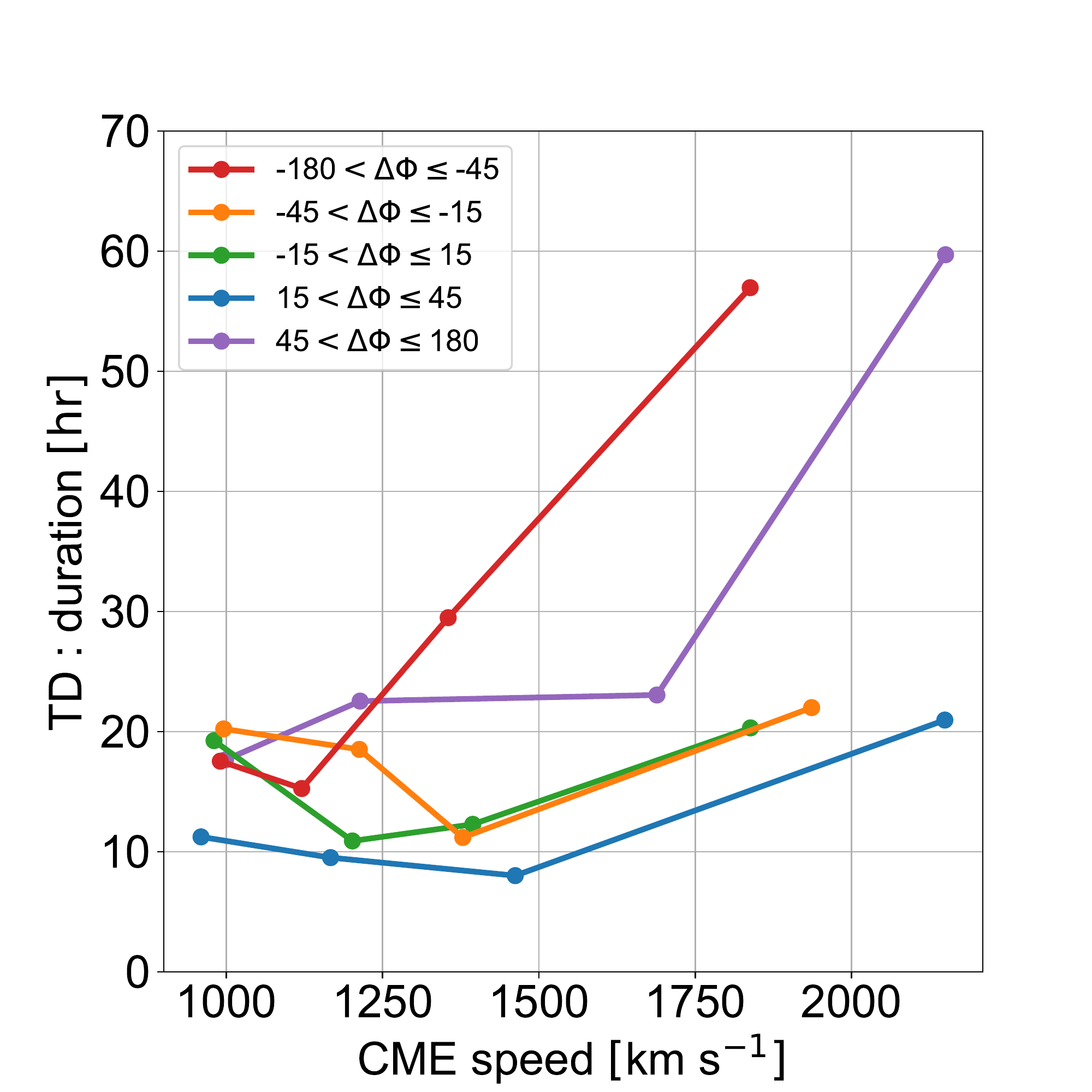}{0.5\textwidth}{(b)}}
\caption{Same as Figure {\ref{TO_speed}}, but for the duration TD.
The number of data points is the same as in Figure {\ref{ts_ps}}(c).
}
\label{TD_speed}
\end{figure*}

\begin{figure*}
\gridline{\fig{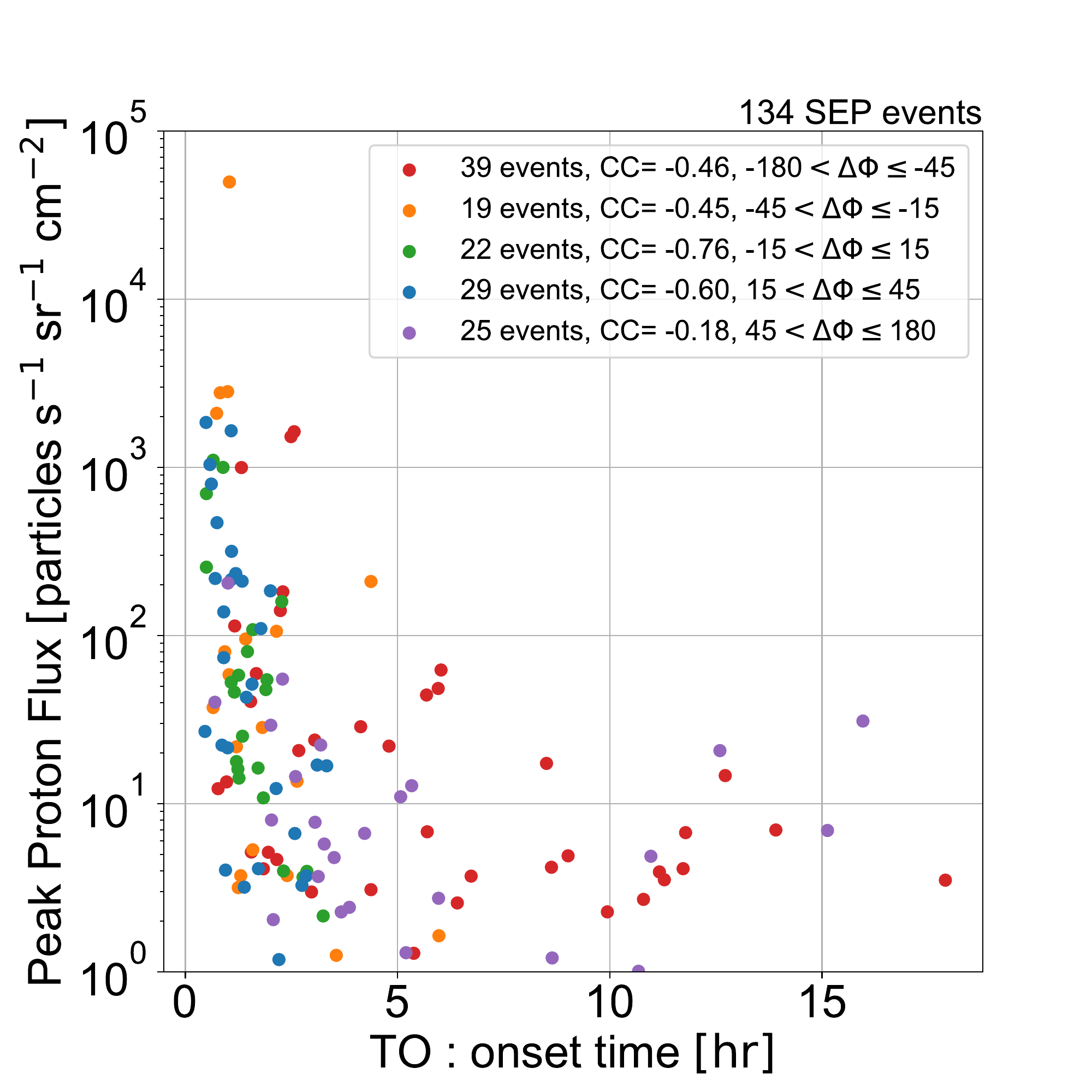}{0.5\textwidth}{}}
\caption{SEP onset time TO vs peak proton flux. The color and used data are same as Figure {\ref{TO_speed}}(a).
The correlation coefficients are calculated between TO and logarithm of peak proton flux.}
\label{TO_Ip}
\end{figure*}

Next we consider how TO, TR, and TD depend on the CME speed ($v_{CME}$), as plotted in 
Figures~{\ref{TO_speed}}, {\ref{TR_speed}}, and {\ref{TD_speed}}, respectively. 
Panel (a) in each of these figures plots individual data, color coded to distinguish 
five ranges of the relative longitude as indicated in the legends.
We produced panel (b) by grouping all the events into five
longitudinal ranges, sorting each group into four subgroups by
$v_{CME}$, and finally taking the average in each of the four $v_{CME}$ subgroups.
This analysis follows the work by \citet{kahler_2013_apj}, and it is intended to make
statistical trends easier to discern.  We adjusted the ranges of $v_{CME}$ in the four subgroups in each longitude range so that each subgroup contains roughly the same number of events.
In Figure {\ref{TD_speed}}, we note a reduced number of events that belong to each of the five longitudinal groups.  This is because for a number of events the SEP half-peak (end) could not be measured, and therefore TD could not be calculated.
According to Figures~{\ref{TO_speed}}\,--\,{\ref{TD_speed}}, the correlations of the timescales with $v_{CME}$ are not strong.  The 
apparent correlations
are susceptible to grouping of the events in different longitude ranges, and taking medians instead of averages does not make the correlations any more solid. Nevertheless, we do see positive correlations for 
TR and TD 
with $v_{CME}$, especially for poorly connected (large $| \Delta \Phi |$) events.  Furthermore, TO appears to be negatively correlated for small  $| \Delta \Phi |$ events.
These trends are consistent with previous results \citep[e.g.,][]{pan_2011_sp, kahler_2013_apj}. 
Finally, the relationship between TO and peak proton flux is shown in Figure~{\ref{TO_Ip}}.
TO is the only timescale that shows a correlation with the peak proton flux.  The correlation is rather strong especially near the footpoint of the Parker spiral.
This trend was reported by \citet{kahler_2013_apj}.

\section{Discussion}

In order to clarify further the roles of CME-driven shocks in generating SEP events, we have carried out a statistical study of the SEP occurrences, peak fluxes, and timescales with respect to the CME source locations and speeds. As noted in Section~1, there are several other factors that could affect the SEP properties, such as preconditioning by earlier events \citep[e.g.,][and references therein]{li_2005_icrc}, spatial distributions of shock parameters closely related to particle acceleration and their time-dependent connections to the observer \citep[e.g.,][and references therein]{kouloumvakos_2019_apj}, transport effects \citep[e.g.,][and references therein]{zhang_2009_apj}, etc.  How these factors as a whole affect the SEP properties can ideally be studied in individual events with a comprehensive approach that involves both data analyses and numerical simulations.  On the other hand, it is also important to know the trend of how the SEP properties vary with respect to the CME source locations and speeds, as found in a large sample of events.
Our approach of starting from CMEs, rather than from SEP events, complements other works that discuss the properties only of CMEs that are associated with SEP events. Recently, \citet{lario_2020_apj} found 11 fast and wide CMEs that did not produce SEP events at any of the 
three locations, 
Earth, STEREO-A, or STEREO-B.  They discussed these events in terms of a deficit in the release of particles at the time of the eruption and the limited extent of the strongest regions of the shocks driven by the CMEs.  In this paper, we presented the statistical trends found in a large number of events observed at GOES, STEREO-A and STEREO-B.  
Moreover, we analyzed mostly solar cycle 24 events, whereas similar statistical studies dealt with solar cycle 23 events \citep[e.g.,][]{kahler_2005_apj, kahler_2013_apj, pan_2011_sp} or partially included solar cycle 24 events \citep[e.g.,][]{papaioannou_2016_jswsc}.
Another difference from previous studies of SEP timescales is that we used the CME source longitude relative to the footpoint of the Parker spiral ($\Delta \Phi$) rather than to the observer. We noted that $\Delta \Phi$ has been used previously for discussing SEP intensities in multi-spacecraft observations \citep[e.g.,][]{richardson_2014_sp}.

We found the source regions of 239 fast ($v_{CME} \geq$ 900~km~s$^{-1}$) and wide (angular width $\geq$ 60$\arcdeg$) CMEs.
Thanks to STEREO data, we were able to determine the CME source regions accurately, even on the far side of the Sun from the Earth, which was not possible for the solar cycle 23 events.  We looked for SEPs at each location, and we calculated the SEP association rate with respect to the CME source longitude relative to the observer (Figures~{\ref{ov_sl}}(c) and {\ref{hz}}). We discussed only $>$10~MeV protons here, partly because they define the space weather effects characterized by NOAA. We found higher SEP association rates for CMEs that originate in what we usually consider to be well-connected longitudes. This suggests that the acceleration of $>$10~MeV protons is more efficient at the nose of a CME-driven shock than elsewhere, which also has been suggested to be the case for higher-energy protons \citep{gopalswamy_2013_apj}. On the other hand, the SEP association rate is non-zero over a wide range of longitudes, consistent with multi-spacecraft observations of SEP events \citep[e.g.,][]{richardson_2014_sp}. For CMEs faster than 900~km~s$^{-1}$, only the narrower range of longitudes E180-E135 may be SEP-free.
In this work, we did not extensively study the effect of the CME width on SEP events except that we limited to CMEs that were wider than 60$\arcdeg$. We, however, note that full halo CMEs that have angular width of 360$\arcdeg$
more often accompany an SEP event than other CMEs (93/136 vs 33/103), supporting the recent study by \citet{lario_2020_apj} that CMEs without SEPs tend to be narrow.  

At the other extreme, a CME from well-connected longitudes must be faster than 1900~km~s$^{-1}$ to be always associated with a SEP event (Figure~\ref{ov_sp}(c)), suggesting the importance of shock waves, even when the CME originates in this longitude range.
Moreover, the peak proton flux from the regression line of well-connected events are one order of magnitude higher than that of poorly connected events with 2000~km~s$^{-1}$, while the peak proton flux of well-connected events vary by three orders of magnitude 
(see Figure~4(a)). It implies that some important factors (e.g., preconditioning by earlier events) other than CME speed and source longitudes exist. 
Another factor that may account for the scatter of the peak SEP flux is the spatial extension of the CME-driven shock wave, which can vary for CMEs with similar speeds.  This has recently been diagnosed using the bandwidth of hectometric type II bursts \citep{iwai_2020_apj}.

By analyzing the timescales of SEP events, we found that TO (the SEP onset time relative to the CME launch) is correlated more tightly with $\Delta\Phi$ (the source longitude relative to the footpoint of the Parker spiral) than is TR (the SEP rise time) for $| \Delta \Phi | <$ 60$\arcdeg$ 
(Figure~6).
The SEP onset is defined by the first-arriving particles. This finding seems to support the idea that only the first-arriving particles may be scatter-free.  Particles that arrive later, even well before the peak, may undergo scattering, possibly by irregularities in the magnetic field, turbulence, etc., in the corona and interplanetary space. Past studies either showed no correlation \citep{kahler_2005_apj,pan_2011_sp} or a 
weak
inverse correlation \citep{kahler_2013_apj} between TO and $v_{CME}$.  We followed \citet{kahler_2013_apj} in analyzing the grouping of $v_{CME}$ into four subgroups and calculating median values---here we showed averages instead of medians.   Our analysis also found weak inverse correlations between TO and $v_{CME}$, as shown by \citet{kahler_2013_apj} for the well-connected longitudes.  
Once protons are accelerated to $>$10~MeV, the strength of the shock, as approximated by the CME speed, may not affect the scatter-free transport of the first-arriving particles. 
As in previous works \citep{pan_2011_sp, kahler_2013_apj}, we found a positive correlation of 
TR and 
TD with $v_{CME}$ (Figures~\ref{TR_speed}(b) and \ref{TD_speed}(b)).
The positive correlation of TR and TD with $v_{CME}$ may be explained if faster CMEs somehow produce wider areas over the shock surface that are favorable for particle acceleration. As a result. the observer would be connected to the regions that accelerate particles for longer time even though the magnetic field connection may change.
We did not use Tm for our analysis because, as discussed by \citet{pan_2011_sp}, the full peak times do not necessarily represent typical peak times for each event, since proton time profiles often show multiple peaks. Moreover, though we did not show the graph, the tendency of Tm was similar to that of TR, so there is no problem in assuming that TR represents a typical peak time.

Concerning the question of how the timescales are related to the SEP flux, we found an inverse correlation between the peak proton flux and TO (but not TR or TD), similar to \citet{kahler_2013_apj}, who pointed out that coarse time bin (half-hour) compromise the timing analysis and instead introduced a background effect to explain the apparent inverse correlation.
We measured timescales in five-minute data, and found a strong negative correlation especially around the footpoint of the Parker spiral (Figure~\ref{TO_Ip}). This could be a consequence of the correlations between CME speed and TO, and CME speed and peak proton flux, but we may speculate that, in SEP events with shorter TO, the observer may connect to the CME-driven shock wave close to the Sun, while it is still strong and efficient in accelerating particles. 

We found that the peak SEP flux appears to be correlated with the magnitude of the solar flare 
(Figure~{\ref{other_cor}}(c)).  This does not necessarily mean that solar flares produce SEPs.  All the flares included in the plot are  associated with fast CMEs, and their magnitude is also weakly correlated with the CME speed (Figure~{\ref{other_cor}}(b)). It is well-established that there are no SEPs from intense flares if there are no CMEs \cite[e.g., flares in AR 12192 in late October 2014; see][]{sun_2015_apj}, and some of the most intense SEP events in solar cycle 23 were associated with flares that were quite modest \citep{cliver_2016_apj}.  However, the GOES soft X-ray flux data are more readily available for space weather operations in real time.  If combined with EUV imagery that shows low coronal signatures for CMEs---such as coronal dimming and post-eruption arcades---the information on solar flares may contribute to SEP forecasting. Another advantage of solar flares is that projection effects may not be as severe as for CMEs in estimating the speed.  It is well known that the true speed of a CME may differ from the projected speed, especially when it is launched far from the limb \citep{burkepile_2004_jgr}.  In this study we used only the speed derived from LASCO observations, which may be an underestimate, especially for halo CMEs. In the future, we plan to compute the 3D speeds of CMEs, using the cone model \citep{xie_2004_jgr} or the graduated cylindrical shell (GCS) model \citep{thernisien_2006_apj}.

\section{Summary and Conclusions}

We conducted a statistical study of the SEP associations of all the fast and wide CMEs that occurred between December 2006 and October 2017.  Our primary findings are summarized as follows:

\begin{enumerate}
\item{The SEP association rate is higher for CMEs that come from the range of longitude of E20\,--\,W100 relative to the observer.}
\item{A CME originating in a well-connected longitude needs to be faster than $\sim$2000~km~s$^{-1}$ to ensure 100\% association with a SEP event.}
\item{The correlation of the peak SXR flux with the peak SEP flux is comparable to that of the CME speed only when the flare is associated with a fast and wide CME.}
\item{The SEP onset time tends to be short when the CME source region is close in longitude (within $\pm$60$\arcdeg$) to the footpoint of the Parker spiral.  This trend is still present but weaker for the SEP rise time.}
\item{There are inverse correlations between the SEP onset timescale and the CME speed in events from regions close in longitude to the footpoint of the Parker spiral.}
\item{There are positive correlations of the SEP rise timescale and duration with the CME speed.}
\item{There are inverse correlations between the peak proton flux and onset timescale.}

\end{enumerate}

In addition to computing the de-projected velocities of CMEs using well-established models, our next steps may include studying proton spectra, associations with electron events and radio bursts, and detailed characterizations of eruptions in the low corona before they arrive at coronagraphic heights.

\acknowledgments

This study is based on the discussion at the Coordinated Data Analysis Workshops held in August 2018 and 2019 held under the auspice of the Project for Solar-Terrestrial Environment Prediction (PSTEP). 
We thank the reviewer for their helpful comments on the manuscript.
This work was supported by MEXT/JSPS KAKENHI Grant Number JP15H05814. The work of N.V.N. was supported by NASA grants 80NSSC18K1126 and 80NSSC20K0287. 
This work was also supported by the joint research project of the Unit
of Synergetic Studies for Space,
Kyoto University and BroadBand Tower, Inc (BBT).
This work benefited from the open data policies of NASA (for SOHO, SDO, STEREO and Wind data) and NOAA (for GOES X-ray and proton data).

\newpage

\startlongtable
\begin{deluxetable*}{lrrlllrrlrrlr}
\label{allcme}
\tabletypesize{\scriptsize}
\tablecaption{Properties of Fast and Wide CMEs and Associated SEP Events}
\tablehead{
\colhead{CME parameters} & \colhead{} & \colhead{} & \colhead{} & \colhead{} & 
\colhead{GOES} & \colhead{} & \colhead{} & \colhead{STEREO-A} & \colhead{} & \colhead{} &
\colhead{STEREO-B} & \colhead{}\\ \cline{1-3} \cline{6-7} \cline{9-10} \cline{12-13}
\colhead{launch} & \colhead{speed} & \colhead{width} & 
\colhead{Flare Class} & \colhead{Flare site} & \colhead{quality} & 
\colhead{Ip} & \colhead{} & \colhead{quality} & \colhead{Ip} & \colhead{} &
\colhead{quality} & \colhead{Ip}\\
\colhead{date,time} & \colhead{(km/s)} & \colhead{(deg)} & 
\colhead{} & \colhead{} & \colhead{} & 
\colhead{(pfu)} & \colhead{} & \colhead{} & \colhead{(pfu)} & \colhead{} & 
\colhead{} & \colhead{(pfu)}
}
\startdata
 2006-12-13 02:25 &   1774 &    360 &  X3.4 &   S06W23 &           Good &    698.0 & &           Good &    592.1 & &           Good &    567.4 \\
 2006-12-14 22:00 &   1042 &    360 &  X1.5 &   S06W46 &           Good &    215.0 & &          HiB\_9 &   [23.6]$^{\mathrm{a}}$ & &          HiB\_7 &   [22.1]$^{\mathrm{a}}$ \\
 2007-01-25 06:33 &   1367 &    360 &  C6.3 &  S08E090 &         No SEP &      --- & &         No SEP &      --- & &         No SEP &      --- \\
 2007-05-19 12:56 &    958 &    106 &  B9.5 &   N07W06 &         No SEP &      --- & &    No SEP$^{\mathrm{b}}$ &      --- & &    No SEP$^{\mathrm{b}}$ &      --- \\
 2007-12-31 00:37 &    995 &    164 &  C8.3 &  S08E081 &         No SEP &      --- & &         No SEP &      --- & &         No SEP &      --- \\
 2008-03-25 18:37 &   1103 &    112 &  M1.7 &  S13E078 &         No SEP &      --- & &         No SEP &      --- & &         No SEP &      --- \\
 2010-02-12 22:13 &   1180 &    120 &  C3.0 &   N25W56 &         No SEP &      --- & &         No SEP &      --- & &         No SEP &      --- \\
 2010-02-13 21:36 &   1005 &     63 &   --- &   N30W63 &         No SEP &      --- & &         No SEP &      --- & &         No SEP &      --- \\
\enddata
\end{deluxetable*}
\tablecomments{
\begin{enumerate}
\item[$^{\mathrm{a}}$] We found a SEP event around the peak even though we missed the onset.
\item[$^{\mathrm{b}}$] The proton flux increased, but did not reach 1~pfu.
\item[$^{\mathrm{c}}$] The peak proton flux may be underestimated due to an energetic storm particle (ESP).
\end{enumerate}
}
\newpage

\startlongtable
\begin{deluxetable*}{lrrlllrrrrr}
\label{allsep}
\tabletypesize{\scriptsize}
\tablecaption{Timescales of SEP Events}
\tablehead{
\colhead{CME parameters} & \colhead{} & \colhead{} & 
\colhead{} & \colhead{} & \colhead{} & \colhead{} &
\colhead{Timescales} & \colhead{} & \colhead{} & \colhead{}\\ \cline{1-3} \cline{8-11}
\colhead{launch} & \colhead{speed} & \colhead{width} & 
\colhead{Flare site} & \colhead{instrument} & \colhead{quality} & \colhead{Ip} &
\colhead{TO} & \colhead{TR} & \colhead{TD} & \colhead{Tm}\\
\colhead{date,time} & \colhead{(km/s)} & \colhead{(deg)} & 
\colhead{} & \colhead{} & \colhead{} & \colhead{(pfu)} & 
\colhead{(min)} & \colhead{(min)} & \colhead{(min)} & \colhead{(min)}
}
\startdata
 2006-12-13 02:25 &   1774 &    360 &   S06W23 &      GOES &           Good &    698.0 &    30.0 &   180.0 &    605 &   420.0 \\
 2006-12-14 22:00 &   1042 &    360 &   S06W46 &      GOES &           Good &    215.0 &    65.0 &    45.0 &     75 &   135.0 \\
 2010-08-14 09:49 &   1205 &    360 &   N17W52 &      GOES &           Good &     14.2 &    76.0 &    50.0 &    280 &   176.0 \\
 2010-08-18 05:29 &   1471 &    184 &  N17W101 &      GOES &           Good &      3.7 &   171.0 &    70.0 &    630 &   321.0 \\
 2011-03-07 19:51 &   2125 &    360 &   N30W48 &      GOES &           Good &     47.7 &   114.0 &   240.0 &   1415 &   539.0 \\
 2011-03-21 02:13 &   1341 &    360 &  N16W129 &      GOES &           Good &      8.0 &   122.0 &   180.0 &   1785 &   522.0 \\
 2011-06-04 21:47 &   2425 &    360 &  N16W153 &      GOES &  Contaminated &      4.9 &   658.0 &   870.0 &    --- &  1818.0 \\
 2011-06-07 06:15 &   1255 &    360 &   S21W54 &      GOES &           Good &     52.7 &    65.0 &   125.0 &   1115 &   245.0 \\
\enddata
\end{deluxetable*}
\tablecomments{
\begin{enumerate}
\item[$^{\mathrm{a}}$] Peak proton flux and timescales may be underestimated due to an energetic storm particle (ESP).
\end{enumerate}
}

\bibliography{bibliography}{}

\begin{thebibliography}{}
\expandafter\ifx\csname natexlab\endcsname\relax\def\natexlab#1{#1}\fi
\providecommand{\url}[1]{\href{#1}{#1}}
\providecommand{\dodoi}[1]{doi:~\href{http://doi.org/#1}{\nolinkurl{#1}}}
\providecommand{\doeprint}[1]{\href{http://ascl.net/#1}{\nolinkurl{http://ascl.net/#1}}}
\providecommand{\doarXiv}[1]{\href{https://arxiv.org/abs/#1}{\nolinkurl{https://arxiv.org/abs/#1}}}

\bibitem[{{Anastasiadis} {et~al.}(2017){Anastasiadis}, {Papaioannou},
  {Sandberg}, {Georgoulis}, {Tziotziou}, {Kouloumvakos}, \&
  {Jiggens}}]{anastasiadis_2017_sp}
{Anastasiadis}, A., {Papaioannou}, A., {Sandberg}, I., {et~al.} 2017, \solphys,
  292, 134, \dodoi{10.1007/s11207-017-1163-7}

\bibitem[{{Brueckner} {et~al.}(1995){Brueckner}, {Howard}, {Koomen},
  {Korendyke}, {Michels}, {Moses}, {Socker}, {Dere}, {Lamy}, {Llebaria},
  {Bout}, {Schwenn}, {Simnett}, {Bedford}, \& {Eyles}}]{Brueckner_1995_sp}
{Brueckner}, G.~E., {Howard}, R.~A., {Koomen}, M.~J., {et~al.} 1995, \solphys,
  162, 357, \dodoi{10.1007/BF00733434}

\bibitem[{{Burkepile} {et~al.}(2004){Burkepile}, {Hundhausen}, {Stanger}, {St.
  Cyr}, \& {Seiden}}]{burkepile_2004_jgr}
{Burkepile}, J.~T., {Hundhausen}, A.~J., {Stanger}, A.~L., {St. Cyr}, O.~C., \&
  {Seiden}, J.~A. 2004, Journal of Geophysical Research (Space Physics), 109,
  A03103, \dodoi{10.1029/2003JA010149}

\bibitem[{{Cane} {et~al.}(2002){Cane}, {Erickson}, \&
  {Prestage}}]{cane_2002_jgr}
{Cane}, H.~V., {Erickson}, W.~C., \& {Prestage}, N.~P. 2002, Journal of
  Geophysical Research (Space Physics), 107, 1315, \dodoi{10.1029/2001JA000320}

\bibitem[{{Cane} {et~al.}(1988){Cane}, {Reames}, \& {von
  Rosenvinge}}]{cane_1988_jgr}
{Cane}, H.~V., {Reames}, D.~V., \& {von Rosenvinge}, T.~T. 1988, \jgr, 93,
  9555, \dodoi{10.1029/JA093iA09p09555}

\bibitem[{{Cliver}(1982)}]{cliver_1982_sp}
{Cliver}, E.~W. 1982, \solphys, 75, 341, \dodoi{10.1007/BF00153481}

\bibitem[{{Cliver}(2016)}]{cliver_2016_apj}
---. 2016, \apj, 832, 128, \dodoi{10.3847/0004-637X/832/2/128}

\bibitem[{{Desai} \& {Giacalone}(2016)}]{desai_2016_lrsp}
{Desai}, M., \& {Giacalone}, J. 2016, Living Reviews in Solar Physics, 13, 3,
  \dodoi{10.1007/s41116-016-0002-5}

\bibitem[{{Dierckxsens} {et~al.}(2015){Dierckxsens}, {Tziotziou}, {Dalla},
  {Patsou}, {Marsh}, {Crosby}, {Malandraki}, \&
  {Tsiropoula}}]{dierckxsens_2015_sp}
{Dierckxsens}, M., {Tziotziou}, K., {Dalla}, S., {et~al.} 2015, \solphys, 290,
  841, \dodoi{10.1007/s11207-014-0641-4}

\bibitem[{{G{\'o}mez-Herrero} {et~al.}(2015){G{\'o}mez-Herrero}, {Dresing},
  {Klassen}, {Heber}, {Lario}, {Agueda}, {Malandraki}, {Blanco},
  {Rodr{\'\i}guez-Pacheco}, \& {Banjac}}]{gomez_herrero_2015_apj}
{G{\'o}mez-Herrero}, R., {Dresing}, N., {Klassen}, A., {et~al.} 2015, \apj,
  799, 55, \dodoi{10.1088/0004-637X/799/1/55}

\bibitem[{{Gopalswamy} {et~al.}(2013){Gopalswamy}, {Xie}, {Akiyama}, {Yashiro},
  {Usoskin}, \& {Davila}}]{gopalswamy_2013_apj}
{Gopalswamy}, N., {Xie}, H., {Akiyama}, S., {et~al.} 2013, \apjl, 765, L30,
  \dodoi{10.1088/2041-8205/765/2/L30}

\bibitem[{{Gopalswamy} {et~al.}(2004){Gopalswamy}, {Yashiro}, {Krucker},
  {Stenborg}, \& {Howard}}]{gopalswamy_2004_jgr}
{Gopalswamy}, N., {Yashiro}, S., {Krucker}, S., {Stenborg}, G., \& {Howard},
  R.~A. 2004, Journal of Geophysical Research (Space Physics), 109, A12105,
  \dodoi{10.1029/2004JA010602}

\bibitem[{{Gopalswamy} {et~al.}(2016){Gopalswamy}, {Yashiro}, {Thakur},
  {M{\"a}kel{\"a}}, {Xie}, \& {Akiyama}}]{gopalswamy_2016_apj}
{Gopalswamy}, N., {Yashiro}, S., {Thakur}, N., {et~al.} 2016, \apj, 833, 216,
  \dodoi{10.3847/1538-4357/833/2/216}

\bibitem[{{Gosling}(1993)}]{gosling_1993_jgr}
{Gosling}, J.~T. 1993, \jgr, 98, 18937, \dodoi{10.1029/93JA01896}

\bibitem[{{Grechnev} {et~al.}(2015){Grechnev}, {Kiselev}, {Meshalkina}, \&
  {Chertok}}]{grechnev_2015_sp}
{Grechnev}, V.~V., {Kiselev}, V.~I., {Meshalkina}, N.~S., \& {Chertok}, I.~M.
  2015, \solphys, 290, 2827, \dodoi{10.1007/s11207-015-0797-6}

\bibitem[{{Guo} {et~al.}(2018){Guo}, {Dumbovi{\'c}}, {Wimmer-Schweingruber},
  {Temmer}, {Lohf}, {Wang}, {Veronig}, {Hassler}, {Mays}, {Zeitlin},
  {Ehresmann}, {Witasse}, {Freiherr von Forstner}, {Heber}, {Holmstr{\"o}m}, \&
  {Posner}}]{guo_2018_sw}
{Guo}, J., {Dumbovi{\'c}}, M., {Wimmer-Schweingruber}, R.~F., {et~al.} 2018,
  Space Weather, 16, 1156, \dodoi{10.1029/2018SW001973}

\bibitem[{{Hudson} \& {Cliver}(2001)}]{hudson_2011_jgr}
{Hudson}, H.~S., \& {Cliver}, E.~W. 2001, \jgr, 106, 25199,
  \dodoi{10.1029/2000JA904026}

\bibitem[{{Iwai} {et~al.}(2020){Iwai}, {Yashiro}, {Nitta}, \&
  {Kubo}}]{iwai_2020_apj}
{Iwai}, K., {Yashiro}, S., {Nitta}, N.~V., \& {Kubo}, Y. 2020, \apj, 888, 50,
  \dodoi{10.3847/1538-4357/ab57ff}

\bibitem[{{Kahler}(1982)}]{kahler_1982_jgr}
{Kahler}, S.~W. 1982, \jgr, 87, 3439, \dodoi{10.1029/JA087iA05p03439}

\bibitem[{{Kahler}(2001)}]{kahler_2001_jgr}
---. 2001, \jgr, 106, 20947, \dodoi{10.1029/2000JA002231}

\bibitem[{{Kahler}(2005)}]{kahler_2005_apj}
---. 2005, \apj, 628, 1014, \dodoi{10.1086/431194}

\bibitem[{{Kahler}(2013)}]{kahler_2013_apj}
---. 2013, \apj, 769, 110, \dodoi{10.1088/0004-637X/769/2/110}

\bibitem[{{Kahler} {et~al.}(1978){Kahler}, {Hildner}, \& {Van
  Hollebeke}}]{kahler_1978_sp}
{Kahler}, S.~W., {Hildner}, E., \& {Van Hollebeke}, M.~A.~I. 1978, \solphys,
  57, 429, \dodoi{10.1007/BF00160116}

\bibitem[{{Kahler} {et~al.}(2001){Kahler}, {Reames}, \&
  {Sheeley}}]{kahler_2001_apj}
{Kahler}, S.~W., {Reames}, D.~V., \& {Sheeley}, N.~R., J. 2001, \apj, 562, 558,
  \dodoi{10.1086/323847}

\bibitem[{{Kahler} {et~al.}(1984){Kahler}, {Sheeley}, {Howard}, {Michels},
  {Koomen}, {McGuire}, {von Rosenvinge}, \& {Reames}}]{kahler_1984_jgr}
{Kahler}, S.~W., {Sheeley}, N.~R., J., {Howard}, R.~A., {et~al.} 1984, \jgr,
  89, 9683, \dodoi{10.1029/JA089iA11p09683}

\bibitem[{{Kahler} \& {Vourlidas}(2005)}]{kahler_2005_jgr}
{Kahler}, S.~W., \& {Vourlidas}, A. 2005, Journal of Geophysical Research
  (Space Physics), 110, A12S01, \dodoi{10.1029/2005JA011073}

\bibitem[{{Kouloumvakos} {et~al.}(2019){Kouloumvakos}, {Rouillard}, {Wu},
  {Vainio}, {Vourlidas}, {Plotnikov}, {Afanasiev}, \&
  {{\"O}nel}}]{kouloumvakos_2019_apj}
{Kouloumvakos}, A., {Rouillard}, A.~P., {Wu}, Y., {et~al.} 2019, \apj, 876, 80,
  \dodoi{10.3847/1538-4357/ab15d7}

\bibitem[{{Lario} {et~al.}(2020){Lario}, {Kwon}, {Balmaceda}, {Richardson},
  {Krupar}, {Thompson}, {Cyr}, {Zhao}, \& {Zhang}}]{lario_2020_apj}
{Lario}, D., {Kwon}, R.~Y., {Balmaceda}, L., {et~al.} 2020, \apj, 889, 92,
  \dodoi{10.3847/1538-4357/ab64e1}

\bibitem[{{Laurenza} {et~al.}(2009){Laurenza}, {Cliver}, {Hewitt}, {Storini},
  {Ling}, {Balch}, \& {Kaiser}}]{laurenza_2009_sw}
{Laurenza}, M., {Cliver}, E.~W., {Hewitt}, J., {et~al.} 2009, Space Weather, 7,
  S04008, \dodoi{10.1029/2007SW000379}

\bibitem[{{Lee} {et~al.}(2012){Lee}, {Mewaldt}, \& {Giacalone}}]{lee_2012_ssr}
{Lee}, M.~A., {Mewaldt}, R.~A., \& {Giacalone}, J. 2012, \ssr, 173, 247,
  \dodoi{10.1007/s11214-012-9932-y}

\bibitem[{{Li} \& {Zank}(2005)}]{li_2005_icrc}
{Li}, G., \& {Zank}, G.~P. 2005, in International Cosmic Ray Conference,
  Vol.~1, 29th International Cosmic Ray Conference (ICRC29), Volume 1, 173

\bibitem[{{Lin} \& {Hudson}(1976)}]{lin_1976_sp}
{Lin}, R.~P., \& {Hudson}, H.~S. 1976, \solphys, 50, 153,
  \dodoi{10.1007/BF00206199}

\bibitem[{{Luhmann} {et~al.}(2008){Luhmann}, {Curtis}, {Schroeder}, {McCauley},
  {Lin}, {Larson}, {Bale}, {Sauvaud}, {Aoustin}, {Mewaldt}, {Cummings},
  {Stone}, {Davis}, {Cook}, {Kecman}, {Wiedenbeck}, {von Rosenvinge}, {Acuna},
  {Reichenthal}, {Shuman}, {Wortman}, {Reames}, {Mueller-Mellin}, {Kunow},
  {Mason}, {Walpole}, {Korth}, {Sanderson}, {Russell}, \&
  {Gosling}}]{luhmann_2008_ssr}
{Luhmann}, J.~G., {Curtis}, D.~W., {Schroeder}, P., {et~al.} 2008, \ssr, 136,
  117, \dodoi{10.1007/s11214-007-9170-x}

\bibitem[{{Mewaldt} {et~al.}(2008){Mewaldt}, {Cohen}, {Cook}, {Cummings},
  {Davis}, {Geier}, {Kecman}, {Klemic}, {Labrador}, {Leske}, {Miyasaka},
  {Nguyen}, {Ogliore}, {Stone}, {Radocinski}, {Wiedenbeck}, {Hawk}, {Shuman},
  {von Rosenvinge}, \& {Wortman}}]{mewaldt_2008_ssr}
{Mewaldt}, R.~A., {Cohen}, C.~M.~S., {Cook}, W.~R., {et~al.} 2008, \ssr, 136,
  285, \dodoi{10.1007/s11214-007-9288-x}

\bibitem[{{Miller} {et~al.}(1997){Miller}, {Cargill}, {Emslie}, {Holman},
  {Dennis}, {LaRosa}, {Winglee}, {Benka}, \& {Tsuneta}}]{miller_1997_jgr}
{Miller}, J.~A., {Cargill}, P.~J., {Emslie}, A.~G., {et~al.} 1997, \jgr, 102,
  14631, \dodoi{10.1029/97JA00976}

\bibitem[{{Nitta} {et~al.}(2014){Nitta}, {Aschwanden}, {Freeland}, {Lemen},
  {W{\"u}lser}, \& {Zarro}}]{nitta_2014_sp}
{Nitta}, N.~V., {Aschwanden}, M.~J., {Freeland}, S.~L., {et~al.} 2014,
  \solphys, 289, 1257, \dodoi{10.1007/s11207-013-0388-3}

\bibitem[{{Onsager} {et~al.}(1996){Onsager}, {Grubb}, {Kunches}, {Matheson},
  {Speich}, {Zwickl}, \& {Sauer}}]{onsager_1996_spie}
{Onsager}, T., {Grubb}, R., {Kunches}, J., {et~al.} 1996, Society of
  Photo-Optical Instrumentation Engineers (SPIE) Conference Series, Vol. 2812,
  {Operational uses of the GOES energetic particle detectors}, ed. E.~R.
  {Washwell}, 281--290, \dodoi{10.1117/12.254075}

\bibitem[{{Pan} {et~al.}(2011){Pan}, {Wang}, {Wang}, \& {Xue}}]{pan_2011_sp}
{Pan}, Z.~H., {Wang}, C.~B., {Wang}, Y., \& {Xue}, X.~H. 2011, \solphys, 270,
  593, \dodoi{10.1007/s11207-011-9763-0}

\bibitem[{{Papaioannou} {et~al.}(2016){Papaioannou}, {Sandberg},
  {Anastasiadis}, {Kouloumvakos}, {Georgoulis}, {Tziotziou}, {Tsiropoula},
  {Jiggens}, \& {Hilgers}}]{papaioannou_2016_jswsc}
{Papaioannou}, A., {Sandberg}, I., {Anastasiadis}, A., {et~al.} 2016, Journal
  of Space Weather and Space Climate, 6, A42, \dodoi{10.1051/swsc/2016035}

\bibitem[{{Pesnell} {et~al.}(2012){Pesnell}, {Thompson}, \&
  {Chamberlin}}]{pesnell_2012_sp}
{Pesnell}, W.~D., {Thompson}, B.~J., \& {Chamberlin}, P.~C. 2012, \solphys,
  275, 3, \dodoi{10.1007/s11207-011-9841-3}

\bibitem[{{Reames}(1999)}]{reames_1999_ssr}
{Reames}, D.~V. 1999, \ssr, 90, 413, \dodoi{10.1023/A:1005105831781}

\bibitem[{{Reames}(2013)}]{reames_2013_ssr}
---. 2013, \ssr, 175, 53, \dodoi{10.1007/s11214-013-9958-9}

\bibitem[{{Richardson} {et~al.}(2014){Richardson}, {von Rosenvinge}, {Cane},
  {Christian}, {Cohen}, {Labrador}, {Leske}, {Mewaldt}, {Wiedenbeck}, \&
  {Stone}}]{richardson_2014_sp}
{Richardson}, I.~G., {von Rosenvinge}, T.~T., {Cane}, H.~V., {et~al.} 2014,
  \solphys, 289, 3059, \dodoi{10.1007/s11207-014-0524-8}

\bibitem[{{Rodriguez} {et~al.}(2017){Rodriguez}, {Sandberg}, {Mewaldt},
  {Daglis}, \& {Jiggens}}]{rodriguez_2017_sw}
{Rodriguez}, J.~V., {Sandberg}, I., {Mewaldt}, R.~A., {Daglis}, I.~A., \&
  {Jiggens}, P. 2017, Space Weather, 15, 290, \dodoi{10.1002/2016SW001533}

\bibitem[{{Smart} \& {Shea}(1996)}]{smart_1996_asr}
{Smart}, D.~F., \& {Shea}, M.~A. 1996, Advances in Space Research, 17, 113,
  \dodoi{10.1016/0273-1177(95)00520-O}

\bibitem[{{Sun} {et~al.}(2015){Sun}, {Bobra}, {Hoeksema}, {Liu}, {Li}, {Shen},
  {Couvidat}, {Norton}, \& {Fisher}}]{sun_2015_apj}
{Sun}, X., {Bobra}, M.~G., {Hoeksema}, J.~T., {et~al.} 2015, \apjl, 804, L28,
  \dodoi{10.1088/2041-8205/804/2/L28}

\bibitem[{{Thernisien} {et~al.}(2006){Thernisien}, {Howard}, \&
  {Vourlidas}}]{thernisien_2006_apj}
{Thernisien}, A.~F.~R., {Howard}, R.~A., \& {Vourlidas}, A. 2006, \apj, 652,
  763, \dodoi{10.1086/508254}

\bibitem[{{Trottet} {et~al.}(2015){Trottet}, {Samwel}, {Klein}, {Dudok de Wit},
  \& {Miteva}}]{trottet_2015_sp}
{Trottet}, G., {Samwel}, S., {Klein}, K.~L., {Dudok de Wit}, T., \& {Miteva},
  R. 2015, \solphys, 290, 819, \dodoi{10.1007/s11207-014-0628-1}

\bibitem[{{Van Hollebeke} {et~al.}(1975){Van Hollebeke}, {Ma Sung}, \&
  {McDonald}}]{vanhol_1975_sp}
{Van Hollebeke}, M.~A.~I., {Ma Sung}, L.~S., \& {McDonald}, F.~B. 1975,
  \solphys, 41, 189, \dodoi{10.1007/BF00152967}

\bibitem[{{von Rosenvinge} {et~al.}(2008){von Rosenvinge}, {Reames}, {Baker},
  {Hawk}, {Nolan}, {Ryan}, {Shuman}, {Wortman}, {Mewaldt}, {Cummings}, {Cook},
  {Labrador}, {Leske}, \& {Wiedenbeck}}]{von_rosenvinge_2008_ssr}
{von Rosenvinge}, T.~T., {Reames}, D.~V., {Baker}, R., {et~al.} 2008, \ssr,
  136, 391, \dodoi{10.1007/s11214-007-9300-5}

\bibitem[{{Xie} {et~al.}(2004){Xie}, {Ofman}, \& {Lawrence}}]{xie_2004_jgr}
{Xie}, H., {Ofman}, L., \& {Lawrence}, G. 2004, Journal of Geophysical Research
  (Space Physics), 109, A03109, \dodoi{10.1029/2003JA010226}

\bibitem[{{Yashiro} {et~al.}(2004){Yashiro}, {Gopalswamy}, {Michalek}, {St.
  Cyr}, {Plunkett}, {Rich}, \& {Howard}}]{yashiro_2004_jgr}
{Yashiro}, S., {Gopalswamy}, N., {Michalek}, G., {et~al.} 2004, Journal of
  Geophysical Research (Space Physics), 109, A07105,
  \dodoi{10.1029/2003JA010282}

\bibitem[{{Zhang} {et~al.}(2007){Zhang}, {Richardson}, {Webb}, {Gopalswamy},
  {Huttunen}, {Kasper}, {Nitta}, {Poomvises}, {Thompson}, {Wu}, {Yashiro}, \&
  {Zhukov}}]{zhang_2007_jgr}
{Zhang}, J., {Richardson}, I.~G., {Webb}, D.~F., {et~al.} 2007, Journal of
  Geophysical Research (Space Physics), 112, A10102,
  \dodoi{10.1029/2007JA012321}

\bibitem[{{Zhang} {et~al.}(2009){Zhang}, {Qin}, \& {Rassoul}}]{zhang_2009_apj}
{Zhang}, M., {Qin}, G., \& {Rassoul}, H. 2009, \apj, 692, 109,
  \dodoi{10.1088/0004-637X/692/1/109}

\end{thebibliography}
\bibliographystyle{aasjournal}

\end{document}